\documentclass[12pt]{article}

\usepackage[dvipdfmx]{graphicx}
\usepackage[hang,small,bf]{caption}
\usepackage[subrefformat=parens]{subcaption}
\usepackage{amsmath}
\usepackage{amsthm}
\usepackage{amssymb}
\usepackage{amscd}
\usepackage{color}

\theoremstyle{definition}
\newtheorem{theorem}{Theorem}[section]
\newtheorem{prop}[theorem]{Proposition}
\newtheorem{lemma}[theorem]{Lemma}
\newtheorem{corollary}[theorem]{Corollary}

\newtheorem{example}[theorem]{Example}
\newtheorem{remark}[theorem]{Remark}

\numberwithin{equation}{section}
\newcommand\bea{\begin{eqnarray}}
\newcommand\ena{\end{eqnarray}}
\newcommand\non{\nonumber}
\newcommand\CV{{\cal V}}
\newcommand\tpartial{\tilde{\partial}}
\newcommand\tb{\tilde{b}}
\newcommand\Proj{{\rm Proj}}
\newcommand\Spec{{\rm Spec}}
\newcommand\ord{\operatorname{ord}}

\allowdisplaybreaks

\title
{Vertex Operators of the KP hierarchy and  Singular Algebraic Curves
  \\
}

\author{
Atsushi Nakayashiki
\footnote{
Department of Mathematics, Tsuda University,
Kodaira, Tokyo 187-8577, Japan,
{\tt atsushi@tsuda.ac.jp}
}
}

\date{
}

\begin{document}

\maketitle

\begin{abstract}
Quasi-periodic solutions of the KP hierarchy acted by vertex operators 
are studied. 
We show, with the aid of the Sato Grassmannian, that solutions thus constructed correspond to torsion free rank one sheaves on some singular algebraic curves
whose normalizations are the non-singular curves corresponding to the seed quasi-periodic solutions.
It means that the action of the vertex operator has an effect of creating singular points on an  algebraic curve.
We further check, by examples, that solutions obtained here can be considered as solitons on quasi-periodic backgrounds, where the soliton matrices are deterimed by parameters in the vertex operators.

\end{abstract}

\section{Introduction}
We revisit the vertex operators of the KP-hierarchy \cite{DJKM}. 
We apply them to quasi-periodic solutions, that is, solutions which are
 expressed by Riemann's theta functions of non-singular algebraic curves \cite{Kr1977}.
The problem we consider here is what kind of solutions we get 
 in this way.  To study this problem we use the Sato Grassmannian \cite{SS}.
 We show that solutions obatined here correspond to certain
 singular algebraic curves whose normalizations are the non-singular curves 
 of the seed quasi-periodic solutions.  It implies two things.
 First is that the action of the vertex operator of the KP-hierarchy 
 has the effect of creating certain singularities on a curve. 
 Second is that the solutions 
 created by vertex operators describe certain limits of quasi-periodic solutions,
 since singular curves may be considered as limits of non-singular curves.
 We check, by computer simulations, that the solutions here represent 
 solitons on the quasi-periodic backgrounds, where the 
soliton matrices can be extracted from the parameters of the vertex operators.
 It implies that wave patterns of quasi-periodic solutions of the KP equation
 contain various shapes of soliton solutions \cite{Kodama2017,Kodama2018}
 as a part.  Recently interactions of solitons and  quasi-periodic solutions attracted much attention in relation with soliton gases \cite{HMP,Kodama2023-1,El2021}.
 It is interesting to study whether the results in this paper can have some
 application to this subject.
 
Now let us explain the results in more detail along the history of researches.
  During the last two decades it is revealed that the shapes of soliton solutions of the KP equation form various wave patterns like web diagrams \cite{CK2008,Kodama2017,Kodama2018}.
  Those wave patterns are related with 
combinatorics of non-negative Grassmannians and cluster algebras \cite{KW2011,KW2014}. Mathematically soliton solutions, in terms of tau function, 
 are those described by linear combinations of exponential functions and are known to be constructed from 
singular algebraic curves of genus 0 \cite{Manin1978,TD1979}. 
Quasi-periodic solutions of the KP  equation are those written by
Riemann's theta function of non-singular 
algebraic curves of positive genus \cite{Kr1977,SW1985}.
 It is expected that quasi-periodic solutions tend to soliton solutions in certain  genus zero limits \cite{Manin1978,TD1979, AG2015}. Therefore it is quite interesting to study the wave patterns of quasi-periodic solutions incorporating the recent development on soliton solutions.
 
 One strategy to study this problem is to take limits of quasi-periodic solutions and make correspondence between quasi-periodic solutions and soliton 
 solutions. However, to carry out this program was not very easy because it is difficult to compute limits of period matrices.  We have avoided this difficulty by using the Sato Grassmannian and have calculated
  limits of quasi-periodic solutions  for several 
  examples \cite{Nak2018,Nak2019,Nak2020,Nak2021}. 
 Recently  there are important progress in computing the limits of quasi-periodic solutions \cite{AG2018,AFMS, Ichikawa2022,Ichikawa2023}. 
 In \cite{Ichikawa2022},  in particular, some kind of limits have been 
 computed for any Riemann surface.
 However it seems difficult to understand how solitonic structure is incorporated 
 in the wave patterns of quasi-periodic solutions from those results.
 
 In this paper we change the direction of study. 
 Instead of studying the limits of quasi-periodic solutions
 we construct solutions corresponding to degenerate
  algebraic curves which are more covenient to see the relation with soliton solutions.
 In course of studying the degeneration of quasi-periodic solutions 
 we found the following formula (Theorem 4.5 of \cite{Nak2020} ):
 \bea
&&
\lim \tau_{g,0}(t)
=C{\rm e}^{-2\sum_{l=1}^\infty \alpha^l t_{2l}}
\non
\\
&&
\times\Bigl(
{\rm e}^{\eta(t,\alpha^{1/2})} \tau_{g-1,0}(t-[\alpha^{-1/2}])
+
(-1)^n{\rm e}^{\eta(t,-\alpha^{1/2})} \tau_{g-1,0}(t-[-\alpha^{-1/2}])
\Bigr),
\non
\ena
where $C$ is a certain constant, $t=(t_1,t_2,t_3,...)$, 
$[\kappa]=(\kappa,\kappa^2/2,\kappa^3/3,...)$, 
$\eta(t,p)=\sum_{j=1}^\infty t_j p^j$, $\tau_{n,0}$ is 
a quasi-periodic solution of the KdV hierarchy, expressed in some standard form,
corresponding to a hyperelliptic curve of genus $n$, lim means pinching
a pair of branching points and $\alpha$ is the parameter correponding to the 
pinched point.  
The term inside the bracket of the right hand side has a very special form.
It is rewritten using the vertex operator introduced in \cite{DJKM}:
\bea
&&
X(p,q)=e^{\eta(t,p)-\eta(t,q)} e^{-\eta(\tpartial,p^{-1})+\eta(\tpartial,q^{-1})},
\non
\\
&&
\tpartial=(\partial_1, \partial_2/2,\partial_3/3,...).
\non
\ena
 In fact, for any function $\tau(t)$, we have
 \bea
 &&
 e^{\eta(t,q)}e^{aX(p,q)}\tau(t-[q^{-1}])=e^{\eta(t,q)}\tau(t-[q^{-1}])
 +ae^{\eta(t,p)}\tau(t-[p^{-1}]).
 \non
 \ena
 If we take $(p,q)=(-\alpha^{1/2}, \alpha^{1/2})$, $a=(-1)^n$ and 
 $\tau(t)=\tau_{g-1,0}$, we recover the above formula. The vertex 
 operator transforms a solution of the KP-hierarchy to another one, that is,
 if $\tau(t)$ is a solution then $e^{aX(p,q)}\tau(t)$ is again a solution for any constant $a$ \cite{DJKM}.
The above fact suggests that the limits of quasi-periodic solutions 
may be described by the action of vertex operators on quasi-periodic solutions.
The aim of this paper is, in a sense, to show that this is actually the case.

Let $\tau_0(t)$ be the quasi-periodic solution of the KP-hierarchy 
constructed from the data
$(C,L_{\Delta,e},p_\infty, z)$, where $C$ is a compact Riemann surface of genus $g>0$, $L_{\Delta,e}$ is a certain holomorphic line bundle
on $C$ of degree $g-1$, $p_\infty$ a point of $C$ and $z$ is a local coordinate 
around $p_\infty$ \cite{Kr1977,SW1985,KNTY1988}.
We apply vertex operators with various parameters successively to $\tau_0(t)$.
More precisely, let $M,N\geq 1$, $p_i,q_j$, $1\leq i\leq N$, $1\leq j\leq M$ distinct complex parameters, $A=(a_{i,j})$ an $M\times N$ matrix. 
We consider the vertex operator of the form
\bea
&&
G=e^{\sum_{i=1}^M\sum_{j=1}^N a_{i,j} X(q_i^{-1},p_j^{-1})}.
\non
\ena
We make a certain shift on $\tau_0(t)$, apply $G$ and multiply it by a constant times exponential function and a get new solution $\tau(t)$ (cf. (\ref{tau-VO})). 
We show that $\tau(t)$ is a solution constructed from the data
$(C',{\cal W}_e, p'_\infty,z)$ where $C'$ is a certain singular algebraic curve
whose normalization is $C$, ${\cal W}_e$ is a certain 
torsion free sheaf of rank one on $C'$ , $p'_\infty$ is a point of $C'$ and $z$ a local coordinate around $p'_\infty$. 
To prove these properties we use the Sato Grassmannian, which we denote by UGM. It is the set of certain subspaces of the vector space $V=\mathbb {C}((z))$ and parametrizes all formal power solutions of the KP-hierarchy. 
We recall here that if $\tau(t)$ is a solution of the KP-hierarchy so is $\tau(-t)$.
W remark that the descriptions of the points of UGM corresponding to $\tau(t)$ and $\tau(-t)$ are not very symmetric. The point corresponding to
$\tau(-t)$ is more suitable to describe the geometry.
By this reason we determine the point $W_e$ of UGM corresponding to 
$\tau(-t)$. This is done by examining the properties of the wave function associated with $\tau(t)$ (cf. (\ref{wavefunction})).

The subspace of $V$ corresponding to $\tau_0(-t)$ is a module over the affine 
coordinate ring $R$ of $C\backslash\{p_\infty\}$. 
Since $W_e$ is a subspace of this space, we consider the stabilizer $R_e$ of $W_e$ in $R$.
Following Mumford \cite{Mum1977} and Mulase\cite{Mul1984,Mul1990} we define 
$C'$and ${\cal W}_e$ as a scheme and a sheaf on it respectively 
using $R_e$ and $W_e$.

Finally it should be mentioned that the relation of the vertex operator and the degeneration of Riemann's theta function has also been observed by Yuji Kodama 
from different point of view \cite{Kodama2023-1,Kodama2023-2}.
It is interesting to investigate the relation of the results of this paper 
and those of Kodama.

The paper is organized as follows.
In section 2 the relation of the KP-hierarchy and the Sato Grassmannian is reviewed.
The main results are formulated and stated in section 3. 
The formula for a quasi-periodic solution, the definition and the properties of the 
vertex operator and the result of the action of the vertex operator on a quasi-periodic solution are given here.
 In section 4 the singular algebraic curve and the sheaf on it associated  with the solution considered in the main theorem are defined and their properties are studied. 
 The example of genus one is studied in detal in section 5. 
 Figures of computer simulation
 by Mathematica are presented here.
The proof of the main theorem is given in section 4 and 5. 
Regarding the readability of the paper
 proofs of assertions in section 4 are given 
in appendix A to E.

\section{KP-hierarchy and Sato Grassmannian}
The KP-hierarchy is the equation for a function $\tau(t)$, $t=(t_1,t_2,t_3,...)$
of the form
\bea
&&
\oint \tau(t-s-[\lambda^{-1}])(t+s+[\lambda^{-1}])e^{-2\eta(s,\lambda)}
\frac{d\lambda}{2\pi i}=0,
\label{KP-hierarchy}
\ena
where $s=(s_1,s_2,s_3,...)$, 
\bea
&&
[\mu]=(\mu,\frac{\mu^2}{2},\frac{\mu^3}{3},...),
\hskip5mm
\eta(t,\lambda)=\sum_{n=1}^\infty t_n \lambda^n,
\non
\ena
and $\oint\cdot \frac{d\lambda}{2\pi i}$ means taking the residue at $\lambda=\infty$. By expanding in $\{s_j\}$ the KP-hierarchy is equivalent 
to the infinite set of differential equations which include the KP equation 
in the bilinear form:
\bea
&&
(D_1^4-3D_2^2-4D_1D_3)\tau\cdot \tau=0,
\label{kp-bilinear}
\ena
where the Hirota derivatives $D_i$ is defined, in general, for a function $f(t)$,
as the Taylor coefficients of the expansion of $f(t+s)f(t-s)$ in $s$:
\bea
&&
f(t+s)f(t-s)=\sum \frac{D^\alpha f\cdot f}{\alpha !}s^\alpha,
\non
\ena
where
$\alpha=(\alpha_1,\alpha_2,...)$, 
$D^\alpha=D_1^{\alpha_1}D_2^{\alpha_2}\cdots$,
$\alpha!=\alpha_1!\alpha_2!\cdots$,
$s^\alpha=s_1^{\alpha_1}s_2^{\alpha_2}\cdots$.
If $x=t_1$, $y=t_2$, $t=t_3$ and $u=2\partial_x^2 \log \tau(t)$,  (\ref{kp-bilinear})
implies the KP equation
\bea
&&
3u_{yy}+(-4u_t+6uu_x+u_{xxx})_x=0.
\label{kp-equation}
\ena

The Sato Grassmannian, which we denote by UGM after Sato, parametrizes all
formal power series solutions of the KP-hierarchy. It is defined as follows.

Let $V={\mathbb C}((z))$ be the space of formal Laurent series in one variable 
$z$, $V_\phi={\mathbb C}[z^{-1}]$ the subspace of polynomials in $z^{-1}$ and $V_0=z{\mathbb C}[[z]]$ the subspace of formal power series vanishing at $z=0$.
Then $V=V_\phi\oplus V_0$. The Sato Grassmannian is the set of subspaces 
of $V$ with the same size as $V_\phi$. 
More precisely let $\pi:V\rightarrow V_\phi$ be the projection. Then UGM is the set of a subspace $U$ such that ${\rm Ker} \pi|_U$ and ${\rm Coker} \pi|_U$ is both
finite dimensional and their dimensions are same.

Here we shall give a criterion for a subspace of $V$ to be a point of UGM.
To this end, for $f(z)\in V$, define the order of $f$ by,
\bea
&&
\ord f=-N \text{ if $f(z)=cz^{N}+O(z^{N+1})$ with $c\neq 0$.}
\non
\ena
It describes the order of a pole at $z=0$ if $N$ is negative.
We set 
\bea
&&
V(n)=\{f\in V \,|\, \ord f\leq n\},
\non
\ena
and, for a subspace $U$ of $V$, set $U(n)=U\cap V(n)$.
Then

\begin{prop}\label{criterion}
A subspace $U$ of $V$ belongs to UGM if and only if 
$\dim U(n)=n+1$ for all sufficiently large $n$.
\end{prop}
\vskip2mm
\noindent
{\it Proof.}
Suppose that $\dim U(n)=n+1$ for $n\geq n_0$ with $n_0\geq 0$.
Then $\dim U(n+1)/U(n)=1$ for $n\geq n_0$. It means that
there exist $f_n\in U$, $n\geq n_0+1$, such that
\bea
&&
f_n=z^{-n}+O(z^{-n+1}),
\hskip5mm
n\geq n_0+1.
\non
\ena
We can take $f_n$, $n\geq n_0+1$, as a part of a basis of $U$.
For the remaining part, since $\dim U(n_0)=n_0+1$,
 there exist $f_{m_i}\in U$, $0\leq i\leq n_0$ such that
 $\ord f_{m_i}=m_i$ with $m_0<\cdots<m_{n_0}$.
By the definition
\bea
&&
r:=\dim {\rm Ker} \pi|_U=\sharp\{j\,|\, m_j<0\}.
\non
\ena
Then 
\bea
&&
\dim {\rm Coker} \pi|_U=n_0+1-(n_0+1-r)=r.
\non
\ena
Thus $U\in UGM$.
\qed

\begin{corollary}\label{criterion-W}
Let $W$ be a subspace of $V$. If there exists an integer $N$ such that $\dim W(n)=n-N$ for all sufficiently large $n$, then $U=z^{N+1}W$ is a point of UGM.
\end{corollary}
\vskip2mm
\noindent
{\it Proof.} Since $U(n)=z^{N+1}W(n+N+1)$ and therefore 
$\dim U(n)=\dim W(n+N+1)=n+1$ for sufficiently large $n$.
Thus the assertion follows from Proposition \ref{criterion}.
\qed

\vskip5mm
\noindent
\begin{example} $\displaystyle W={\mathbb C}z^3+{\mathbb C}z^2+\sum_{i=4}^\infty {\mathbb C}z^{-i}$. For $n\geq 4$  $\dim W(n)=n-1$. 
Then $\displaystyle U=z^2W={\mathbb C}z^5+{\mathbb C}z^4+
\sum_{i=2}^\infty {\mathbb C}z^{-i}$ is a point of UGM
since ${\rm Ker}\pi|_U={\mathbb C}z^5+{\mathbb C}z^4$ and 
${\rm Coker}\pi|_U={\mathbb C}1+{\mathbb C}z^{-1}$.
\end{example}

Next we explain the correspondence between solutions of the KP-hierarchy and 
points of UGM. For a point of UGM the corresponding solution of the KP-hierarchy is constructed as a series using Schur functions and Pl\"ucker coordinates. For this see \cite{Nak2019}.

We need, in this paper, the converse construction of the point of UGM from 
a solution of the KP-hierarchy. Let $\tau(t)$ be a solution of the KP-hierarchy.
Define the wave function $\Psi(t;z)$ and the adjoint wave function $\Psi^\ast(t;z)$ by
\bea
&&
\Psi(t;z)=\frac{\tau(t-[z])}{\tau(t)}e^{\eta(t,z^{-1})},
\hskip10mm
\Psi^{\ast}(t;z)=\frac{\tau(t+[z])}{\tau(t)}e^{-\eta(t,z^{-1})}.
\label{wavefunction}
\ena

\begin{theorem}\label{reverse-U} \cite{SS,KNTY1988}
Let $U$ be the vector space spanned by the expansion coefficients of 
$\tau(t)\Psi^\ast(t;z)$ in $t$. Then $U$ is the point of UGM corresponding to 
$\tau(t)$.
\end{theorem}

It is easy to verify that  if  $\tau(t)$ is a solution of the KP-hierarchy 
so is $\tau(-t)$.

\begin{corollary}\label{reverse-U'}
Let $U'$ be the vector space spanned by the expansion coefficients of 
$\tau(t)\Psi(t;z)$ in $t$ Then $U'$ is the point of UGM corresponding 
to $\tau(-t)$. 
\end{corollary}
\vskip2mm
\noindent
{\it Proof.}  Let $\Psi_{-}^{\ast}(t;z)$ be the adjoint wave function of $\tau(-t)$.
Then
\bea
&&
\tau(-t)\Psi_{-}^{\ast}(t;z)=\tau(-t-[z])e^{-\eta(t,z^{-1})}.
\label{expansion-psi-minus}
\ena
If we set $s=-t$, it is equal to
\bea
&&
\tau(s-[z])e^{\eta(s,z^{-1})}=\tau(s)\Psi(s;z).
\label{expansion-psi}
\ena
  Since the vector space generated by
the expansion coefficients of (\ref{expansion-psi}) in $s$ is the same as that generated 
by expansion coefficients of (\ref{expansion-psi-minus}) in $t$, the assertion of the lemma follows from Theorem \ref{reverse-U}.
\qed

\section{Main results}
Let $C$ be a compact Riemann surface of genus $g>0$, $\{\alpha_i, \beta_i\}_{i=1}^g$ a canonical basis of $H^1(C, {\mathbb Z})$, $\{dv_i\}_{i=1}^g$ the normalized basis of holomorphic one forms,  $\Omega=(\Omega_{i,j})_{1\leq i, j\leq g}$ 
with $\Omega_{i,j}=\int_{\beta_j} dv_i$ the period matrix,  
$J(C)={\mathbb C}^g/L_\Omega$ with $L_\Omega={\mathbb Z}^g+\Omega {\mathbb Z}^g$ the Jacobian variety of $C$,
$p_\infty$ a point of $C$, $I(p)=\int_{p_\infty}^{p} dv$ with $dv={}^t(dv_1,...,dv_g)$ the Abel map, $K$ Riemann's constant, $\theta(z|\Omega)$ Riemann's 
theta function 
\bea
&&
\theta(z|\Omega)=
\sum_{n\in {\mathbb Z}^g}
\exp(\pi i{}^t n \Omega n+2\pi i {}^t n z),
\quad
z={}^t(z_1,...,z_g),
\non
\ena
where $n\in {\mathbb Z}^g$ is considered as a column vector.

We extend the the definition of the Abel map to divisors of  any
degree by, for $\displaystyle D=\sum_{j=1}^m p_j-\sum_{j=1}^n q_j$,
\bea
&&
I(D)=\sum_{j=1}^m I(p_j)-\sum_{j=1}^n I(q_j).
\non
\ena
We denote by $\Delta$ the Riemann divisor. It is the dvisor of degree $g-1$ which satisfies
$2\Delta\equiv\Omega_C^1$ and $I(\Delta)=K$, where $\equiv$ signifies the linear equivalence of divisors and $\Omega^1_C$ is the linear equivalence class of divisors of holomorphic one forms.  The Riemann divisor is uniquely determined
from the canonical homology basis by the condition \cite{Mum1983-1}
\bea 
&&
\{I(p_1+\cdots+p_{g-1}-\Delta) \,\vert\, p_1,...,p_{g-1}\in C\}=
\{z\in J(C)\, \vert\, \theta(z|\Omega)=0\}.
\non
\ena
Notice that the left hand side does not depend on the choice of the base 
point $p_\infty$ of the Abel map. 

Let $E(p_1,p_2)$ be the prime form \cite{Fay1973, Mum1983} (see also \cite{Nak2016}):
\bea
&&
E(p_1,p_2)=\frac{\theta[\delta](\int_{p_1}^{p_2}dv)}{h_\delta(p_1)h_\delta(p_2)},
\non
\ena
where $\delta=\binom{\delta'}{\delta''}$, $\delta',\delta''\in \frac{1}{2}{\mathbb Z}^g$ is a non-singular odd half characteristic and $h_\delta(p)$ is the half differential satisfying
\bea
&&
h_\delta^2(p)=\sum_{j=1}^g\frac{\partial\theta[\delta]}{\partial z_j}(0)dv_j(p).
\non
\ena

 Take a local coordinate $z$ around $p_\infty$ and write
\bea
&&
E(P_1,P_2)=\frac{E(z_1,z_2)}{\sqrt{dz_1}\sqrt{dz_2}}, 
\hskip5mm P_i\in C, \quad z_i=z(P_i),
\non
\\
&&
d_{z_1}d_{z_2}\log E(z_1,z_2)
=\left(\frac{1}{(z_1-z_2)^2}
+\sum_{i,j=1}^\infty q_{i,j}z_1^{i-1}z_2^{j-1}\right)dz_1dz_2,
\label{logE-exp}
\\
&&
dv_i=\sum_{j=1}^\infty v_{i,j} z^{j-1} dz.
\non
\ena

We set
\bea
&&
\CV=(v_{i,j})_{1\leq i\leq g, 1\leq j},
\hskip10mm 
q(t)=\sum_{i,j=1}^\infty q_{i,j}t_i t_j.
\non
\ena

Then 
\bea
&&
\tau_0(t)=e^{\frac{1}{2}q(t)}\theta(\CV t+e|\Omega)
\non
\ena
is a solution of the KP-hierarchy for arbitrary $e\in {\mathbb C}^g$ \cite{SW1985}(see also \cite{Shiota1986,KNTY1988,Nak2016}).

Let 
\bea
&&
X(p,q)=e^{\eta(t,p)-\eta(t,q)} e^{-\eta(\tpartial,p^{-1})+\eta(\tpartial,q^{-1})},
\non
\\
&&
\eta(t,p)=\sum_{j=1}^\infty t_j p^j,
\hskip10mm
\tpartial=(\partial_1, \partial_2/2,\partial_3/3,...),
\non
\ena
be the vertex operator.

The following theorem is known.

\begin{theorem}{\rm \cite{DJKM}}
If $\tau(t)$ is a solution of the KP-hierarchy, so is $e^{aX(p,q)}\tau(t)$ for any
$a\in {\mathbb C}$.
\end{theorem}

Vertex operators satisfy
\bea
&&
X(p_1,q_1)X(p_2,q_2)=\frac{(p_1-p_2)(q_1-q_2)}{(p_1-q_2)(q_1-p_2)}
:X(p_1,q_1)X(p_2,q_2):
\label{VO-CR}
\ena
where $:\quad :$ denotes the normal ordering taking all differential 
operators to the right of all multiplication operators, that is, 
\bea
&&
:X(p_1,q_1)X(p_2,q_2):=
e^{\sum_{j=1}^2(\eta(t,p_j)-\eta(t,q_j))} e^{\sum_{j=1}^2(-\eta(\tpartial,p_j^{-1})+\eta(\tpartial,q_j^{-1}))}.
\non
\ena
The following properties follow from this.
\bea
X(p_1,q_1)X(p_2,q_2)&=&X(p_2,q_2)X(p_1,q_1) 
\quad\text{if  $p_1\neq q_2$ and $q_1\neq p_2$,}
\non
\\
X(p_1,q_1)X(p_2,q_2)&=&0
\quad \text{if $p_1=p_2$ or $q_1=q_2$}.
\non
\ena

Let $M,N$ be positive integers, $q_i$, $1\leq i\leq M$, $p_j$, $1\leq j\leq N$ non-zero complex numbers and  $(a_{i,j})$ an $M\times N$ complex matrix. 
We set $p_{N+j}=q_j$ and use both notation $p_{N+j}$ and $q_j$.

Set
\bea
&&
G=e^{\sum_{i=1}^M\sum_{j=1}^N a_{i,j} X(q_i^{-1},p_j^{-1})}.
\non
\ena
Define
\bea
&&
\tau(t)=\Delta(p_1^{-1},...,p_N^{-1})e^{\sum_{j=1}^N\eta(t,p_j^{-1})}
G \, \tau_0(t-\sum_{j=1}^N[p_j]),
\label{tau-VO}
\ena
where $\Delta(p_1,...,p_n)=\prod_{1\leq i<j\leq n}(p_j-p_i)$. It can be computed explicitly as follows.

Set $L=M+N$ and define the $L\times N$ matrix $B=(b_{i,j})$ by
\bea
b_{i,j}&=&\delta_{i,j} \quad \text{for $1\leq i,j\leq N$},
\non
\\
b_{N+i,j}&=&a_{i,j}\prod_{m\neq j}^N\frac{p_j^{-1}-p_m^{-1}}{q_i^{-1}-p_m^{-1}}
 \quad \text{for $1\leq i\leq M$, $1\leq j\leq N$},
\label{matrix-B}
\ena
that is,
\bea
 &&
 B=\left(
 \begin{array}{ccc}
 1&\quad&\quad\\
 \quad&\ddots&\quad\\
 \quad&\quad&1\\
 b_{N+1,1}&\cdots&b_{N+1,N}\\
 \vdots&\quad&\vdots\\
 b_{N+M,1}&\cdots&b_{N+M,N}\\
 \end{array}
 \right).
 \non
 \ena
We set $[L]=\{1,...,L\}$ and denote by $\binom{[L]}{N}$ the set of $(i_1,...,i_N)$, $1\leq i_1<\cdots<i_N\leq L$ \cite{Kodama2017,Kodama2018}. 
For $I=(i_1,...,i_N)\in \binom{[L]}{N}$ set
\bea
&& 
\Delta^{-}_I=\Delta(p_{i_1}^{-1},...,p_{i_N}^{-1}),
\hskip5mm
\eta_I=\sum_{i\in I} \eta(t,p_i^{-1}),
\hskip5mm
[p_I]=\sum_{i\in I} [p_i],
\non
\\
&&
B_I=\det(b_{i_r,s})_{1\leq r,s\leq N}.
\non
\ena

By a direct calculation using the commutation relation (\ref{VO-CR}) we have 

\begin{prop}\cite{Nak2021}
The function $\tau(t)$ of (\ref{tau-VO}) has the following expression
\bea
&&
\tau(t)=\sum_{I\in \binom{[L]}{N}} B_I \Delta^{-}_I e^{\eta_I}\tau_0(t-[p_I]).
\label{tau-explicit}
\ena
\end{prop}

\begin{remark}\label{positivity-general} The matrices $(a_{i,j})$ and $B$ correspond to each other.
When we study the positivity of  $\tau(t)$, it is convenient to begin 
with the matrix $B$ and define the matrix $(a_{i,j})$ from it. 
For example, in the case where $\tau_0(t)>0$ for any real $t$,
 $\{p_i\}$ are real and $p_1^{-1}<\cdots<p_L^{-1}$, 
then $\tau(t)>0$ if $B_I\geq 0$ for any $I$.
Later in section 5 this view point is used.
\end{remark}

The part $\tau_0(t-[p_I])$ can further be expressed by theta function.
To write it let $d\tilde{r}_k$, $k\geq 1$, be the normalized differential 
of the second kind with a pole only at $p_\infty$ of order $k+1$, that is,
it satisfies
\bea
&&
\int_{\alpha_j} d\tilde{r}_k=0 \text{ for any $j$}, 
\hskip5mm
d\tilde{r}_k=d\left(z^{-k}-O(z)\right) \text{ near $p_\infty$}.
\non
\ena
The expansion of $d\tilde{r}_k$ near $p_\infty$ can be written more explicitly using $\{q_{i,j}\}$.
In integral form it is given by
\bea
&&
\int^p d\tilde{r}_k=z^{-k}-\sum_{j=1}^\infty q_{k,j}\frac{z^j}{j},
\hskip5mm p\in C, z=z(p).
\label{second-k}
\ena

Then

\begin{prop}\label{tau-theta}
In terms of the theta function $\tau(t)$ defined by (\ref{tau-VO})
is written as
\bea
&&
\tau(t)=e^{\frac{1}{2}q(t)}\sum_{J\in \binom{[L]}{N}} B_J C_J 
e^{\sum_{j\in J}\sum_{k=1}^\infty t_k \int^{P_j} d\tilde{r}_k}
\,\,\theta\left({\cal V}t-\sum_{j\in J} I(Q_j)+e\right),
\label{tau-theta-formula}
\ena
where
\bea
&&
C_J=\prod_{i<j, i,j\in J}E(p_j,p_i)\prod_{j\in J}\frac{p_j}{E(0,p_j)^N},
\non
\ena
and $P_j, Q_j\in C$ such that $z(P_j)=p_j$, $z(Q_j)=q_j$.
\end{prop}

This proposition is proved by a direct calculation using the following lemma which can be derived from (\ref{logE-exp}) .

\begin{lemma}\label{exp-Q}
Let $\displaystyle Q(t|s)=\sum_{i,j=1}^\infty q_{i,j}t_i s_j$. Then
\bea
&&
e^{Q([z]|[w])}=\frac{E(z,w)}{w-z}\frac{zw}{E(0,z)E(0,w)},
\quad
q^{\frac{1}{2}Q([z]|[z])}=\frac{z}{E(0,z)}.
\non
\ena
\end{lemma}

For $e={}^t(e_1,...,e_g)\in {\mathbb C}^g$ $L_e$ denotes the holomorphic line bundle of degree $0$ on $C$ whose characteristic homomorphism is specified by
\bea
&&
\chi(\alpha_j)=1, 
\hskip10mm
\chi(\beta_j)=e^{2\pi i e_j}.
\non
\ena
If $c\in {\mathbb C}^g$ is taken such that $\theta(c)\theta(e+c)\neq 0$  then 
$\theta(I(p)+e+c)/\theta(I(p)+c)$ is a meromorphic section of $L_{-e}$.

We denote $L_\Delta$ the holomorphic line bundle of degree $g-1$ corresponding to $\Delta$.  Then we consider $L_{\Delta,-e}:=L_\Delta\otimes L_{-e}$
which has $\theta(I(p)+e)/E(p,p_\infty)$ as a meromorphic section if $\theta(e)\neq 0$.

Let $H^0(C, L_{\Delta,-e}(\ast p_\infty))$ be the vector space of 
meromorphic sections of $ L_{\Delta,-e}$ which are holomorphic 
on $C\backslash\{p_\infty\}$. Using the local coordinate $z$ we embed this 
space into $V={\mathbb C}((z))$ as follows. We consider a section of 
$L_{\Delta,-e}$ as $E(p,p_\infty)^{-1}$ times a multi-valued 
meromorphic function on $C$ whose transformation rule is the same as that of 
$\theta(I(p)+e)$, that is, 
\bea
&&
f(p+\alpha_j)=f(p),
\hskip5mm
f(p+\beta_j)=e^{-\pi i \Omega_{j,j}-2\pi i (\int_{p_\infty}^p dv_j+e_j)}f(p).
\non
\ena
We realize a section of $ L_{\Delta,-e}$ using a function on $C$ in this 
way. Then we expand elements of $H^0(C,  L_{\Delta,-e}(\ast p_\infty))$
around $p_\infty$ 
in $z$ as $\sum a_n z^n\sqrt{dz}$ 
and get elements $\sum a_nz^n$ of $V$. In the following we always consider 
$H^0(C,  L_{\Delta,-e}(\ast p_\infty))$ as a subspace of $V$ in this way.

\begin{theorem}\label{tau-0}\cite{KNTY1988}
The point of UGM corresponding to $\tau_0(t)$ is $zH^0(C,  L_{\Delta,-e}(\ast p_\infty))$ and that corresponding to $\tau_0(-t)$ is $zH^0(C,  L_{\Delta,e}(\ast p_\infty))$.
\end{theorem}

For the solution $\tau(t)$ of (\ref{tau-VO}) the descriptions of the points of UGM correponding to $\tau(t)$ and $\tau(-t)$ are not very symmetric as opposed to 
$\tau_0(t)$. We consider $\tau(-t)$ here, since it is more conveniently related with the geometry of $C$ as in the case of soliton solutions \cite{TD1979,Manin1978}.

Set 
\bea
&&
b'_{N+j,i}=b_{N+j,i} (p_i^{-1}q_j),
\non
\\
&&
W_{e}=\{ f\in H^0(C,  L_{\Delta,e}(\ast p_\infty))\,\vert\,
f(p_i)=-\sum_{j=1}^M b'_{N+j,i}f(q_j), \quad 1\leq i\leq N\},
\non
\\
&&
U_{e}=z^{N+1}W_{e}.
\ena

Our main theorem is 

\begin{theorem}\label{main}
(i) The subspace $U_{e}$ is a point of UGM.
\vskip2mm
\noindent
(ii) The point of UGM corresponding to $\tau(-t)$ is $U_e$.
\end{theorem}

\section{Singular curve created by vertex operators}
In this section we study the geometry of $U_e$.

By Theorem \ref{tau-0} the point of UGM corresponding to the solution 
$\tau_0(-t)$ is associated with $(C,L_{\Delta,e},p_\infty,z)$, where, as in the previous section, $C$ is a non-singular algebraic curve of genus $g>0$, 
$L_{\Delta,e}$ is the holomorphic line bundle on $C$, $p_\infty$ a point of $C$ and $z$ a local coordinate around $p_\infty$ \cite{SW1985,KNTY1988}.

We show that the point $U_e$ of UGM corresponding to the solution $\tau(-t)$ 
is associated with $(C',{\cal W}_e, p_\infty',z)$, where $C'$ is a singular algebraic 
curve whose normalization is $C$, ${\cal W}_e$ is a rank one torsion free sheaf 
on $C'$, $p_\infty'$ is a point of $C'$ and $z$ a local coordinate at $p_\infty'$.
 It suggests that, geometrically, the action of the vertex operator
has an effect of creating  some kind of singularities on a curve. 
Moreover singular curves may be considered as degenerate limits of non-singular
curves. Therefore $\tau(-t)$ and consequently $\tau(t)$ can be considered as a certain limit of  a quasi-periodic solution.

 In order to define the curve $C'$ from $U_e$ the most appropriate way 
 in the present case 
is the abstract algebraic method of Mumford and Mulase \cite{Mum1977,Mul1984,Mul1990}. 
Namely we define $C'$ as a complete integral scheme and ${\cal W}_e$ as a sheaf on it.

We referred to \cite{Hartshorne,Iitaka, Kawamata, Matsumura} as references on algebraic geometry and commutative algebras.

 Let $R:=H^0(C,{\cal O}(\ast p_\infty))$ be the vector space of meromorphic 
 functions on $C$ which have a pole only at $p_\infty$.
By expanding functions in the local coordinate $z$ around $p_\infty$ 
we consider $R$ as a subspace of $V={\mathbb C}((z))$. 
It is  the affine coordinate ring of $C\backslash\{p_\infty\}$.
The vector space $H^0(C, L_{\Delta,e}(\ast p_\infty))$  is
 an $R$-module and $W_e$ is a vector subspace of it.
Let
\bea
&&
R_e=\{f\in R\,|\, f W_e\subset W_e\}
\non
\ena
be the stabilizer of $W_e$ in $R$. Then

\begin{prop}\label{stabilizer}
We have
\bea
&&
R_e=\{f\in R\,|\, f(p_i)=f(q_j) \text{ if $b_{N+j,i}\neq 0$}\}.
\label{Re}
\ena
\end{prop}
\vskip5mm

The proof of this proposition is given in Appendix A.

To study the structure of $R_e$ we introduce a directed graph 
$G_B$ associated with the matrix $B$. 
The vertices of $G_B$ consists of $\{p_1,...,p_L\}$. 
The vertices $p_i$ and $p_{N+j}$ are connected by an edge with 
the weight $b_{N+j,i}$. The direction of the edge is from $p_{N+j}$ to $p_i$.
We understand  the edge with the weight $0$ is the same as that 
there is no edge. Other edges are not connected. 
Notice that one can recover the matrix $B$
from $G_B$.

 Let $s$ be the number of connected components of $G_B$. We divide
the set of vertices $\{p_i|1\leq i\leq L\}$ according as connected components and rename them as
$\{p_{i,j}|1\leq j\leq n_i\}$,$1\leq i\leq s$. We denote $P_{i,j}$ the point on $C$
such that $z(P_{i,j})=p_{i,j}$.

\vskip5mm
\noindent
\begin{example} Consider
\bea
&&
B=\left(
\begin{array}{cc}
1&0\\
0&1\\
0&a\\
-b&0\\
\end{array}
\right),
\qquad
a,b\neq 0.
\non
\ena
\vskip1cm
In this case $G_B$ is
\vskip2mm

\setlength{\unitlength}{1mm}
\begin{picture}(100,20)(0,0)
\put(40,15){\circle{2}}
\put(33,14){$p_1$}
\put(50,15){\circle{2}}
\put(55,14){$p_4$}
\put(49,15){\vector(-1,0){8}}
\put(43,17){$-b$}

\put(40,5){\circle{2}}
\put(33,4){$p_2$}
\put(50,5){\circle{2}}
\put(55,4){$p_3$}
\put(49,5){\vector(-1,0){8}}
\put(44,7){$a$}

\end{picture}

and $s=2$. Then $(p_{1,1},p_{1,2})=(p_1,p_4)$, $(p_{2,1},p_{2,2})=(p_2,p_3)$ for 
example.
\end{example}

\begin{example}
Let
\bea
&&
B=\left(
\begin{array}{cc}
1&0\\
0&1\\
-c&a\\
-d&b\\
\end{array}
\right),
\qquad
a,b,c,d\neq 0.
\non
\ena
\vskip1cm
Then $G_B$ is

\setlength{\unitlength}{1mm}
\begin{picture}(100,20)(0,0)
\put(40,15){\circle{2}}
\put(33,14){$p_1$}
\put(50,15){\circle{2}}
\put(55,14){$p_3$}
\put(49,15){\vector(-1,0){8}}
\put(43,17){$-c$}

\put(40,5){\circle{2}}
\put(33,4){$p_2$}
\put(50,5){\circle{2}}
\put(55,4){$p_4$}
\put(49,5){\vector(-1,0){8}}
\put(49.5,14){\vector(-1,-1){8.5}}
\put(49,6){\vector(-1,1){8}}
\put(40,15){\circle{2}}
\put(43,1){$-d$}
\put(48,11){$a$}
\put(41,9.5){$b$}
\end{picture}

In this case $s=1$ and $(p_{1,1},p_{1,2},p_{1,3},p_{1,4})=(p_1,p_2,p_3,p_4)$ for 
example.

\end{example}

In the notation introduced above $R_e$ is described as
\bea
&&
R_e=\{f\in R| f(p_{i,j})=f(p_{i,j'}) \text{ for any $j, j'$, $1\leq i\leq s$}\}.
\label{description-Re}
\ena

Let $H^0\left(C,{\cal O}(np_\infty)\right)$ be the space of meromorphic functions 
on $C$ with a pole only at $p_\infty$ of order at most $n$ and 
\bea
&&
R(n)=H^0\left(C,{\cal O}(np_\infty)\right),
\non
\\
&&
R_e(n)=R(n)\cap R_e.
\ena
Notice that $R(n)=R_e(n)=\{0\}$ for $n<0$ and $R(0)=R_e(0)={\mathbb C}$.
The set of subspaces $\{R_e(n)\}$ satisfies  $R_e(n)\subset R_e(n+1)$ for any $n$ and $\displaystyle R_e=\cup_{n=0}^\infty R_e(n)$.

Set
\bea
&&
A'=\oplus_{n=0}^\infty R_e(n),
\non
\\
&&
C'=\Proj\, A',
\non
\ena
We call $R_e(n)$ the homogeneous component of $A'$ with degree $n$.
There is a natural injective morphism
$\varphi:\Spec R_e \rightarrow C'$  
given by
\bea
&&
\varphi({\cal P})= \oplus_{n=0}^\infty {\cal P}^{(n)},
\hskip10mm
{\cal P}^{(n)}={\cal P}\cap R_e(n).
\label{spec-proj}
\ena

Next define
\bea
&&
p'_\infty=\oplus_{n=0}^\infty R_e(n-1),
\non
\ena
where $R_e(n-1)$ is located at the homogeneous 
component of $A'$ with degree $n$.
It can be easily checked that  $p'_\infty\in C'$.

By the Riemann-Roch theorem there exists $N_0$ such that
\bea
&&
\dim R_e(n)/R_e(n-1)=1 \quad n\geq N_0.
\non
\ena
Take an arbitrary $m\geq N_0$ and $a\in R_e(m)$ such that
\bea
&&
a=z^{-m}+O(z^{-m+1}).
\label{element-a}
\ena
We consider $a$ as a homogeneous element of $A'$ with degree $m$.
Set
\bea
D'_+(a)&=&\{{\cal P}\in \Proj\,A'\,|\, a\notin {\cal P}\},
\non
\\
A'_{(a)}&=&\{ua^{-n}\,|\, u\in R_e(mn), \,\,n\geq 0\}
\non
\\
&=&\text{the set of elements of degree zero in $A'[a^{-1}]$}.
\non
\ena
Then $D'_+(a)$ is an affine open subscheme of $C'$ isomorphic to $\Spec A'_{(a)}$(c.f. \cite{Hartshorne}). This isomorphism is given by
\bea
&&
{\cal P} \mapsto \oplus_{n=0}^{\infty}a^{-n}{\cal P}^{(nm)}.
\non
\ena

Then 

\begin{theorem}\label{C'-structure}
(i) $p'_\infty \notin \varphi(\Spec\, R_e)$.
\vskip2mm
\noindent
(ii) $\displaystyle C'=\varphi(\Spec\, R_e) \cup \{p'_\infty\}.$
\vskip2mm
\noindent
(iii) $H^0(C'\backslash\{p'_\infty\}, {\cal O}_{C'})=R_e$.
\vskip2mm
\noindent
(iv) $p'_\infty\in D'_+(a)$ and it corresponds to a maximal ideal of $A'_{(a)}$.
\end{theorem}

The proof of this theorem is given in Appendix B.

By this theorem
\bea
&&
C'=\varphi(\Spec\, R_e)\cup D'_{+}(a)
\non
\ena
is an affine open cover of $C'$. 
The rings $R_e$ and $A'_{(a)}$ are integral domains, since they are subrings 
of ${\mathbb C}((z))$. Moreover $\dim C'=1$ because, as we shall show in
Lemma \ref{K'}, the quotient field of $R_e$ is isomorphic to the quotient field of $R$ which is the field of meromorphic functions on $C$. 
Using Proposition \ref{structure-R-Re} and the Riemann-Roch theorem we can easily prove that $A'$ is generated over ${\mathbb C}$ by a finite number of homogeneous elements. Then $A'$ can be written as a quotient of polynomial ring by a homogeneous ideal. Therefore $C'$ becomes a closed subscheme of a weighted projective space and consequently of a projective space.
Thus $C'$ is a projective integral scheme of dimension one, that is, $C'$ 
is a projective integral curve.

Next we define a sheaf on $C'$.

Let $H^0(C,L_{\Delta,e}(np_\infty))$ be the space of meromorphic sections of 
$L_{\Delta,e}$ on $C$ with a pole only at $p_\infty$ of order at most $n$ and 
\bea
&&
W_e(n)=W_e\cap H^0(C,L_{\Delta,e}(np_\infty)).
\non
\ena
Define
\bea
&&
W^{gr}_e=\oplus_{n=0}^\infty W_e(n).
\non
\ena
Using Lemma \ref{lemma-2} and the Riemann-Roch theorem 
it can easily be proved that $W^{gr}_e$ is a finitely generated $A'$-module.
Therefore  $W^{gr}_e$ defines a coherent ${\cal O}_{C'}$ 
module ${\cal W}_e$ on $C'$ such that
\bea
&&
H^0(C'\backslash\{p_\infty'\}, {\cal W}_e)=W_e.
\non
\ena
Moreover we can prove

\begin{prop}\label{torsion-free}
 The ${\cal O}_{C'}$-module ${\cal W}_e$ is torsion free and of rank one.
\end{prop}

The proof of this proposition is given in Appendix C.

Next we study the relation between $C$ and $C'$.

A compact Riemann surface can be embedded into a projective space. Therefore there is a projective scheme corresponding to $C$ which we denote by the same symbol $C$. 
In terms of $\displaystyle R=\cup_{n=0}^\infty R(n)$ $C$ is described 
as 
\bea
&&
C=\Proj A,
\hskip10mm
A=\oplus_{n=0}^\infty R(n).
\non
\ena

There is an injective morphism, $\varphi$: $\Spec R\rightarrow C$ given by 
a similar formula to (\ref{spec-proj}) which we denote by the same symbol. The affine scheme $\Spec A$ corresponds to $C\backslash\{p_\infty\}$.
Similarly to $p'_\infty$ define
\bea
&&
\tilde{p}_\infty=\oplus_{n=0}^\infty R(n-1),
\non
\ena
where $R(n-1)$ is situated at the degree $n$ component of $A$  as in the previous case.

Set 
\bea
D_+(a)&=&\{{\cal P}\in \Proj\,A\,|\, a\notin {\cal P}\},
\non
\\
A_{(a)}&=&\{ua^{-n}\,|\, u\in R(mn), \,\,n\geq 0\}.
\non
\ena

As in the case of $C'$ the following proposition holds.
\begin{prop}
(i) $\tilde{p}_\infty \notin \varphi(\Spec\, R)$.
\vskip2mm
\noindent
(ii) $\displaystyle C=\varphi(\Spec\, R) \cup \{\tilde{p}_\infty\}.$
\vskip2mm
\noindent
(iii) $\displaystyle H^0(C\backslash\{\tilde{p}_\infty\},{\cal O}_C)=R$.
\vskip2mm
\noindent
(iv) $\tilde{p}_\infty\in D_+(a)$ and it corresponds to a maximal ideal of $A_{(a)}$.
\end{prop}

All elements of the maximal ideal $\tilde{p}_\infty$ of $A_{(a)}$ vanish
at $p_\infty$. So $\tilde{p}_\infty$ can be identified with $p_\infty$.

The inclusion map $A'\subset$ $A$ induces a morphism $\psi:C\rightarrow C'$. 
Then 

\begin{prop}\label{normalization}
The morphism $\psi:C\rightarrow C'$ gives the normalization of $C'$.
\end{prop}

The proof of this proposition is given in Appendix D.
\vskip2mm

Finally we study the singularities of $C'$.

Let $P$ be a point of the compact Riemann surface $C$ such that $P\neq p_\infty$, $z(P)=p$ and $m_P\in \Spec R$ the maximal ideal corresponding 
to $P$, that is,
\bea
&&
m_P=\{f\in R| f(p)=0\}.
\non
\ena
Then $m'=\psi(m_P)=m_P\cap R_e$ is a maximal ideal of $R_e$ since $R$ is integral over $R_e$ by Corollary \ref{R-finite-Re}. 
We denote by $R_{m_P}$ the localization of $R$ at $m_P$ etc.
Then 

\begin{prop}\label{singularity}
(i) If $P\neq P_i$ for any $i$,then  $(R_e)_{m'}\simeq R_{m_P}$.
 In particular $(R_e)_{m'}$ is a normal ring and the closed point $m'\in \Spec R_e$
 is a non-singular point.
 \vskip2mm
 \noindent
 (ii) If $P=P_{i,j}$ and $n_i\geq 2$,  then $(R_e)_{m'}$ is not a normal ring. In particular  $m'\in \Spec R_e$ is a singular point. Moreover 
 $\psi^{-1}(m')=\{P_{i,1},...,P_{i,n_i}\}$ in this case.
\end{prop}

The proof of this proposition is given in Appendix E.

This proposition shows that $C'$ is obtained from $C$ by identifying
the points $P_{i,1},...,P_{i,n_i}$ for each $i$ such that $n_i>1$.

\section{Solitons on elliptic backgrounds}
As remarked in remark \ref{positivity-general} if all $t_i$, $p_j$ are real, $\tau_0(t)>0$ for any $t$\cite{DN1989}, $B_I\geq 0$ for any $I$ and 
$p_1^{-1}<\cdots<p_L^{-1}$, then $\tau(t)$ given by (\ref{tau-explicit}) is positive.
Notice that $\tau(t)$ is a linear combination of $\{e^{\eta_I}\tau_0(t-[p_I])\}$.
Then, in the region of the $xy$-plane such that $e^{\eta_I}\tau_0(t-[p_I])$
is dominant, $u(t)=2\partial_x^2\log \tau(t)\approx 2\partial_x^2\log \tau_0(t)$
is the quasi-periodic wave corresponding to the shift of $\tau_0(t)$. On the boundary of 
two such domains, soliton like waves will appear as in the case of  soliton solutions\cite{Kodama2017,Kodama2018}. 
Thus it is expected that $u(t)$ represents a soliton on a quasi-periodic background. In this section we verify it by a comupter simulation in the case of genus one .

Let $a,b$ be positive real numbers. Set $2\omega_1=-ib$, $2\omega_2=ab$, $\Omega=\omega_2/\omega_1=ia$ and 
${\mathbb L}=2\omega_1{\mathbb Z}+2\omega_1{\mathbb Z}$. Define 
\bea
&&
g_2=60\,{\sum}_{\omega\in {\mathbb L}, \omega\neq 0} 
\,\frac{1}{\omega^4},
\qquad
g_3=140\,{\sum}_{\omega\in {\mathbb L}, \omega\neq 0}\,\frac{1}{\omega^6}.
\non
\ena
Then $g_2$, $g_3$ are real.
Consider the algebraic curve $C$ defined by the corresponding Weierstrass cubic
\bea
&&
y^2=4x^3-g_2x-g_3.
\non
\ena
Let $\wp(u)$ be the Weierstrass elliptic function. Then $u \mapsto (\wp(u),\wp'(u))$
gives an isomorphism between the complex torus 
$\mathbb{C}/{\mathbb L}$ and $C$, where $u=0$ corresponds to $\infty\in C$.  A basis of holomorphic one forms is $du=dx/y$. A canonical 
homology basis can be taken such that 
\bea
&&
\int_\alpha du=2\omega_1,
\hskip5mm
\int_\beta du=2\omega_2.
\non
\ena
Therefore the normalized holomorphic one form is given by
\bea
&&
dv=(2\omega_1)^{-1}du.
\non
\ena
We take $u$ as a local coordinate around $\infty$. 
Then the corresponding solution of the KP hierarchy is given by
\bea
&&
\tau_0(t)=e^{\frac{1}{2}q(t)}\theta\left(\frac{x}{2\omega_1}+e|\Omega\right),
\hskip10mm
\label{elliptic-tau}
\ena
where $e$ is an arbitrary complex constant. If we take $e\in i{\mathbb R}$ and 
$x$, $t_j$, $j\geq 2$ to be real then $\tau_0(t)$ is real and positive.

In the present case $q_{i,j}$ defining $q(t)$ is described in the following way.
Let 
\bea
&&
\theta_{11}(z|\Omega)=
\sum_{n\in\mathbb{Z}}e^{\pi i \Omega(n+\frac{1}{2})^2+2\pi i(n+\frac{1}{2})(z+\frac{1}{2})}.
\non
\ena
Sometimes $\theta_{11}(z|\Omega)$ is simply denoted by $\theta_{11}(z)$.
It satisfies
\bea
&&
\theta_{11}(-z)=-\theta_{11}(z), 
\hskip5mm
\theta_{11}(z)=\theta'_{11}(0) z+O(z^3), 
\quad
\theta_{11}'(0)\neq 0.
\non
\ena
The prime form is written as
\bea
&&
E(z_1,z_2)=\frac{2\omega_1}{\theta_{11}'(0)}
\theta_{11}\left(\frac{z_2-z_1}{2\omega_1}\right).
\non
\ena
Therefore 
\bea
&&
\frac{\partial^2}{\partial z_1\partial z_2}
\log \frac{\theta_{11}\left(\frac{z_2-z_1}{2\omega_1}\right)}{z_2-z_1}
=
\sum_{i,j=1}^\infty q_{i,j}z_1^{i-1} z_2^{j-1}.
\label{bilinear-differential}
\ena

Let us take the elliptic solution (\ref{elliptic-tau}) as $\tau_0(t)$ in the general formula (\ref{tau-explicit}).

To neatly write $\tau(t)$ let us introduce
\bea
&&
F(z)=\log \theta_{11}\left(\frac{z}{2\omega_1}\right),
\hskip10mm
F_j(z)=\frac{(-1)^{j-1}}{(j-1)!}F^{(j)}(z).
\non
\ena

By Proposition \ref{tau-theta} we have 

\begin{prop}
For $\tau_0(t)$ given by (\ref{elliptic-tau}) the $\tau(t)$ defined by
(\ref{tau-VO}) is written as
\bea
\tau(t)&=&ce^{\sum_{j=1}^\infty c_j t_j +\frac{1}{2}q(t)}
\non
\\
&&
\times
\sum_{I} B_IC_I
e^{\sum_{j\in I}\sum_{k=1}^\infty t_k F_k(p_j)}
\theta\left(\frac{x-\sum_{j\in I} p_j}{2\omega_1}+e|\Omega\right).
\label{elliptic-general}
\ena
Here $c$, $c_j$ are certain constants and 
\bea
&&
C_I=
\prod_{j\in I}
\frac{p_j}{\theta_{11}\left(\frac{p_j}{2\omega_1}\right)^N}
\prod_{j,k\in I, j<k}
\theta_{11}\left(\frac{p_k-p_j}{2\omega_1}\right).
\non
\ena
\end{prop}

We take $\{t_j\}$, $\{p_j\}$ are real and $e\in i{\mathbb R}$.
If $B_IC_I\geq 0$ for any $I$ and some of them is positive then
the $\sum_I$ part in the right hand side of  (\ref{elliptic-general}) is positive and $u=2\partial^2_x\log \tau(t)$ is non-singular.

For the positivity of $C_I$ we have 

\begin{prop}
If $p_1<\cdots<p_L$, $p_L-p_1<ab$ and $|p_j|<ab$ for $1\leq j\leq L$ then $C_I>0$ for any
$I$.
\end{prop}

This proposition follows from the following lemma which can easily be proved.
\begin{lemma}
If $a>0$, then $i\theta_{11}(ix|ia)>0$ for $0<x<a$.
\end{lemma}
\vskip10mm

\noindent
\begin{example}\label{example1}
Take $M=N=2$, $a=1$, $b=6$, $(p_1,p_2,p_3,p_4)=(0.31,0.46,0.81,4.89)$, 
$t_j=0$ for $j\geq 4$ and
\bea
&&
B=\left(
\begin{array}{cc}
1&0\\
0&1\\
-2&1\\
-3&0\\
\end{array}
\right).
\ena
This $B$ corresponds to (3) of \S4.6.4 \cite{Kodama2018}.
Notice that the matrix in \cite{Kodama2018} is the transpose of our matrix $B$.
In this case $B_{12}=1$, $B_{13}=1$, $B_{14}=0$, $B_{23}=2$, $B_{24}=3$, $B_{34}=3$ and 
\bea
&&
(F^{(1)}(p_1),F^{(1)}(p_2),F^{(1)}(p_3),F^{(1)}(p_4))=(3.25,2.21,1.30,0.05).
\label{soliton-shape}
\ena
The results of a computer simulation of $u(t)$ is given in figures 1,2,3.
We can see the soliton corresponding to the matrix $B$ on the periodic waves.
\end{example}
\vskip5mm

\noindent
\begin{example}\label{example2}
Take $M,N,a,b,p_j, t_j$ the same as those in Example \ref{example1}.  Consider $B$ of the form
\bea
&&
B=\left(
\begin{array}{cc}
1&0\\
0&1\\
-1&2\\
-1&1\\
\end{array}
\right).
\ena
For this matrix $B_{12}=1$, $B_{13}=2$, $B_{14}=1$, $B_{23}=1$, $B_{24}=1$, $B_{34}=1$. This $B$ corresponds to (1) of \S4.6.4 \cite{Kodama2018}.
See figures 4,5,6.
\end{example}

\clearpage

\begin{figure}
\centering

\begin{minipage}{0.4\columnwidth}
\centering
\includegraphics[width=4cm]{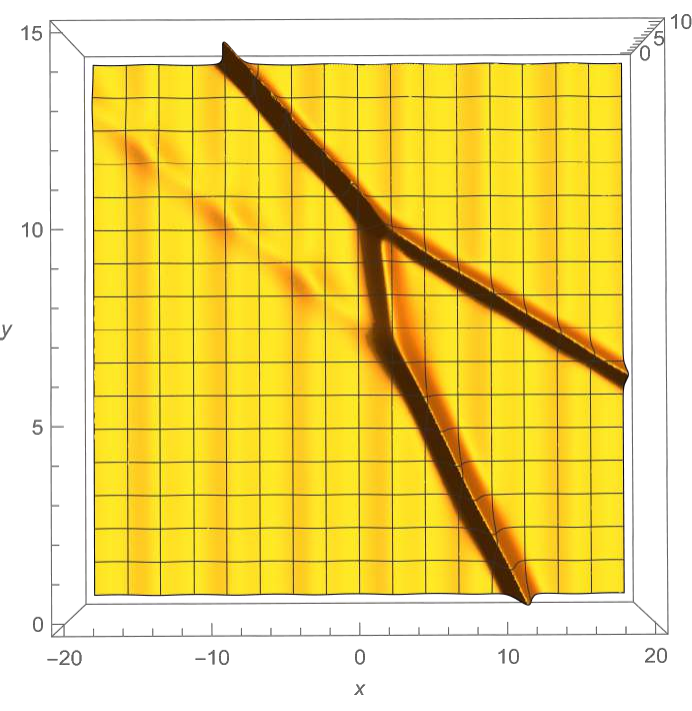}
\subcaption{from above}
\end{minipage}
\hskip1cm
\begin{minipage}{0.4\columnwidth}
\centering
\includegraphics[width=4cm, height=4cm]{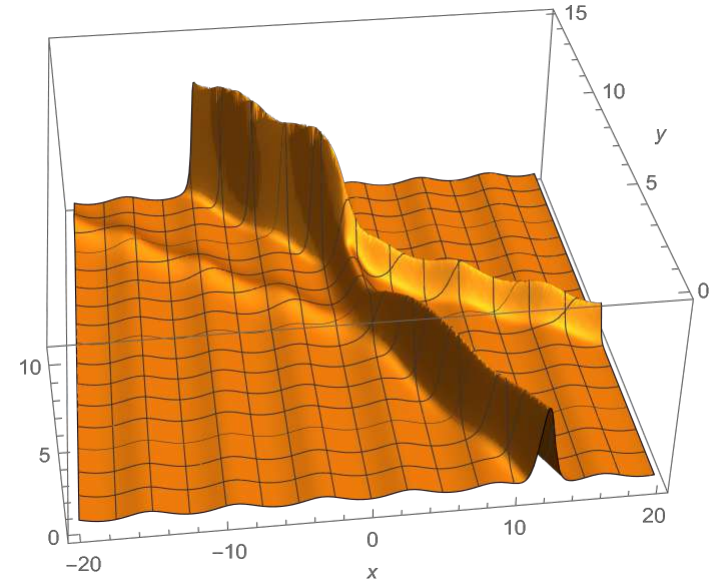}
\subcaption{from an agngle}
\end{minipage}
\caption{Example \ref{example1} $t=-3$, $-20\leq x\leq 20$, $0\leq y\leq 15$}

\end{figure}

\begin{figure}
\centering

\begin{minipage}{0.4\columnwidth}
\centering
\includegraphics[width=4cm]{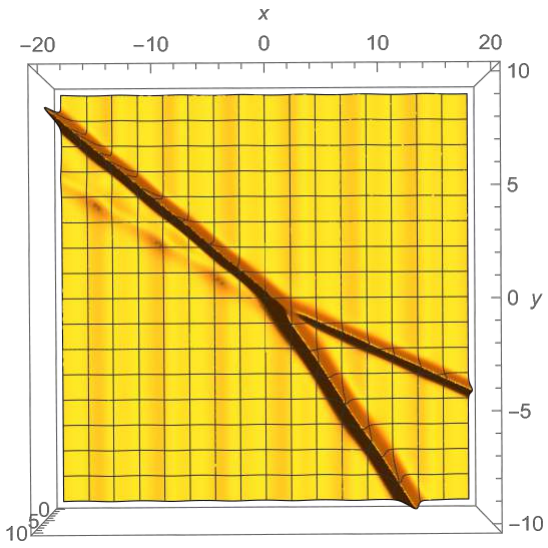}
\subcaption{from above}
\end{minipage}
\hskip1cm
\begin{minipage}{0.4\columnwidth}
\centering
\includegraphics[width=4cm, height=4cm]{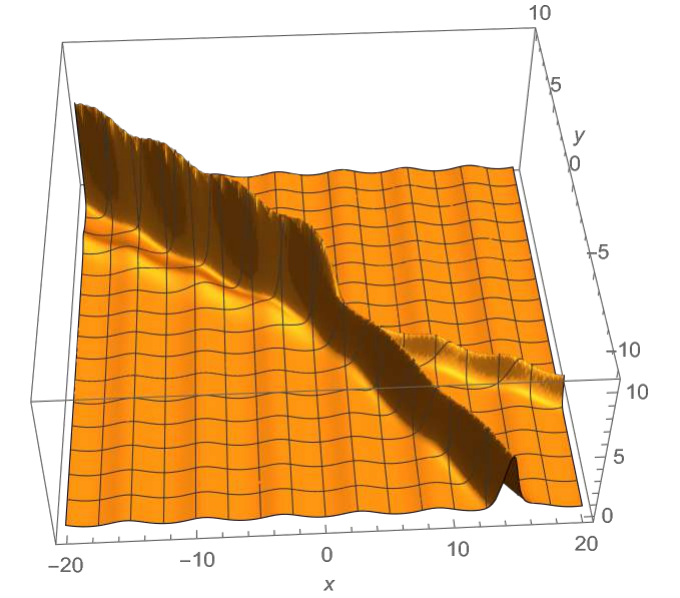}
\subcaption{from an agngle}
\end{minipage}
\caption{Example \ref{example1} $t=0$, $-20\leq x\leq 20$, $-10\leq y\leq 10$}

\end{figure}

\begin{figure}
\centering

\begin{minipage}{0.4\columnwidth}
\centering
\includegraphics[width=4cm]{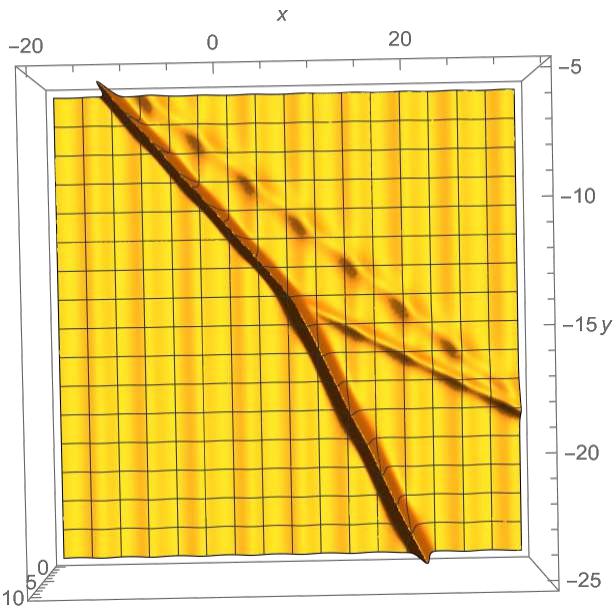}
\subcaption{from above}
\end{minipage}
\hskip1cm
\begin{minipage}{0.4\columnwidth}
\centering
\includegraphics[width=4cm, height=4cm]{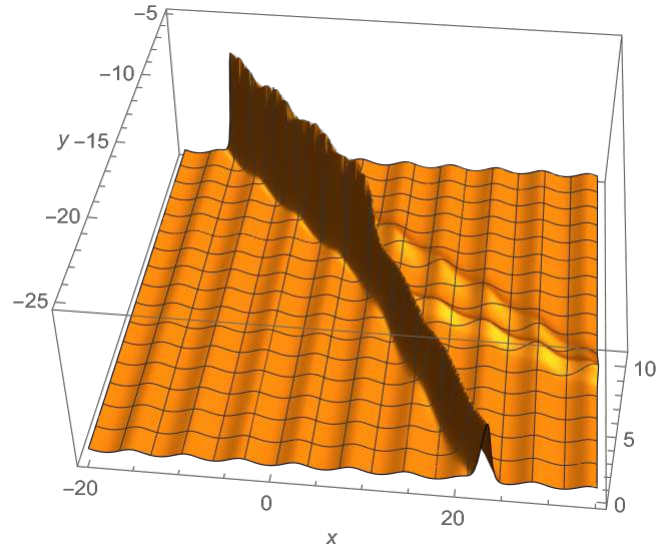}
\subcaption{from an agngle}
\end{minipage}
\caption{Example \ref{example1} $t=3$, $-20\leq x\leq 35$, $-25\leq y\leq -5$}

\end{figure}

\clearpage

\begin{figure}
\centering

\begin{minipage}{0.4\columnwidth}
\centering
\includegraphics[width=4cm]{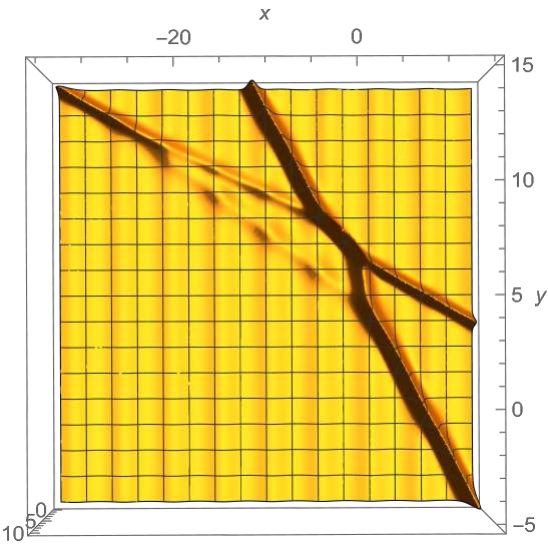}
\subcaption{from above}
\end{minipage}
\hskip1cm
\begin{minipage}{0.4\columnwidth}
\centering
\includegraphics[width=4cm, height=4cm]{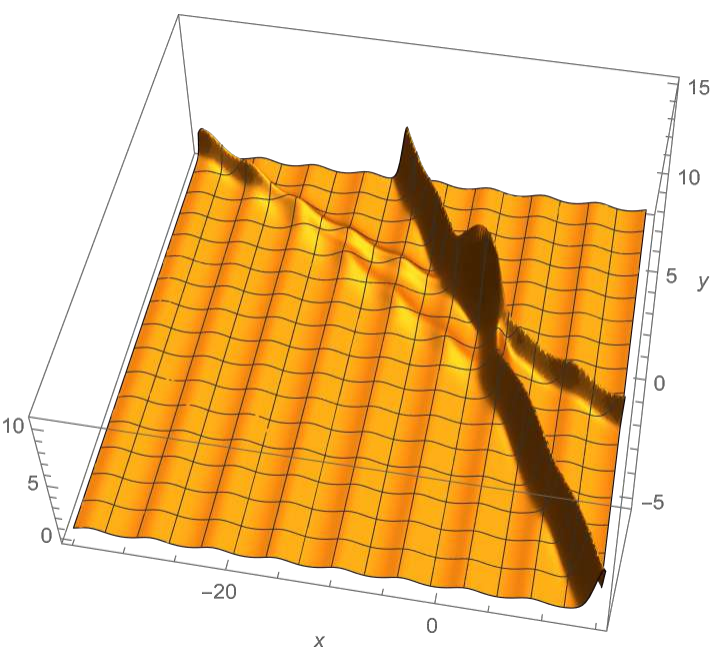}
\subcaption{from an agngle}
\end{minipage}
\caption{Example \ref{example2} $t=-2$, $-35\leq x\leq 15$, $-5\leq y\leq 15$}

\end{figure}

\begin{figure}
\centering

\begin{minipage}{0.4\columnwidth}
\centering
\includegraphics[width=4cm]{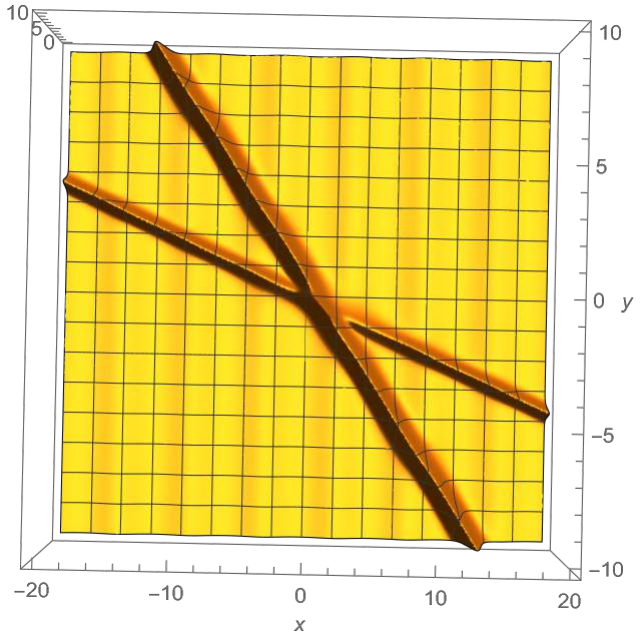}
\subcaption{from above}
\end{minipage}
\hskip1cm
\begin{minipage}{0.4\columnwidth}
\centering
\includegraphics[width=4cm, height=4cm]{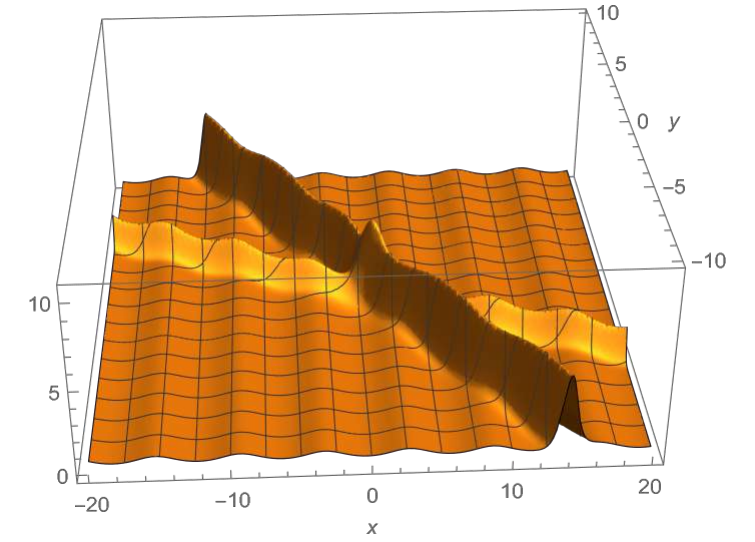}
\subcaption{from an agngle}
\end{minipage}
\caption{Example \ref{example2} $t=0$, $-20\leq x\leq 20$, $-10\leq y\leq 10$}

\end{figure}

\begin{figure}
\centering

\begin{minipage}{0.4\columnwidth}
\centering
\includegraphics[width=4cm]{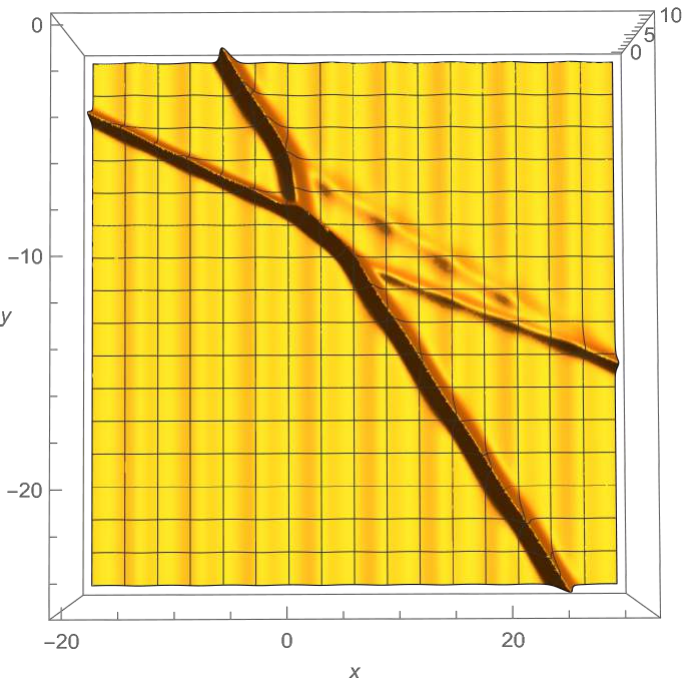}
\subcaption{from above}
\end{minipage}
\hskip1cm
\begin{minipage}{0.4\columnwidth}
\centering
\includegraphics[width=4cm, height=4cm]{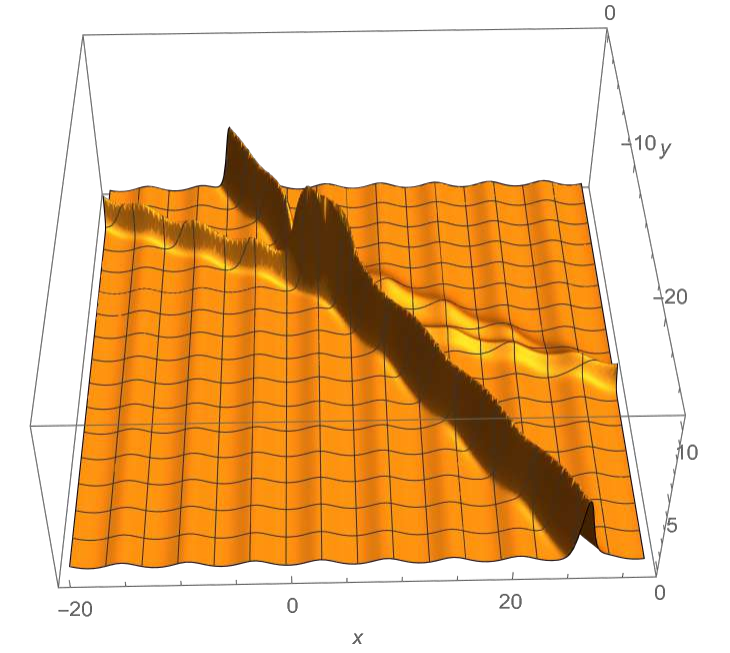}
\subcaption{from an agngle}
\end{minipage}
\caption{Example \ref{example2} $t=2$, $-20\leq x\leq 32$, $-25\leq y\leq 0$}

\end{figure}

\clearpage

\section{Proof of Theorem \ref{main} (i)}
In this section we freely use notation on sheaf cohomologies.
Namely, for a holomorphic line bundle ${\cal L}$, distinct points $S_i$, $1\leq i\leq m$, $S_j'$, $1\leq j\leq m'$ on $C$, $P$ and positive integers $\{m_i\}$,
$\{m_j'\}$ we denote 
\bea
&&
H^0(C,{\cal L}(-\sum m_i S_i+\sum m'_j S_j'+\ast P))
\non
\ena
the space of meromorphic sections of ${\cal L}$  which have a zero at $S_i$ of 
order at least $m_i$, a pole at $S_j$ of order at most $m_j'$ and a pole at $P$ 
of any order.

To prove Theorem \ref{main} (i) it is sufficient to show
\bea
&&
\dim W_e(n)=n-N  \text{ for $ n>>0$},
\label{tobeproved}
\ena
by Corollary \ref{criterion-W}.

\begin{lemma}\label{lemma-1}
There exists $F_i$ in $H^0(C, L_{\Delta,e}(\ast p_\infty))$, $1\leq i\leq L$ such that
\bea
&&
F_i(p_j)=\delta_{i,j}.
\non
\ena
\end{lemma}

The proof of this lemma is given in the end of this section.

For $1\leq m\leq M$ set
\bea
&&
\varphi_m=F_{N+m}-\sum_{i=1}^Nb'_{N+m,i}F_i.
\label{phi-m}
\ena
Then
\bea
&&
\varphi_m\in W_e,
\non
\ena
since
\bea
&&
\varphi_m(q_l)=\delta_{m,l},
\hskip10mm
\varphi_m(p_i)=-b'_{N+m,i},
\label{phi-m-value}
\ena
and 
\bea
&&
-\sum_{j=1}^M b'_{N+j,i}\, \varphi_m(q_j)=-b'_{N+m,i}=\varphi_m(p_i).
\non
\ena

Notice that the vector space 
\bea
&&
H^0(C,L_{\Delta,e}(-\sum_{j=1}^L P_j+\ast p_\infty))
\non
\ena
  is a subspace of $W_e$, since elements of it vanish at all $p_j$ and the linear equations imposed in $W_e$ are trivially satisfied.

\begin{lemma}\label{lemma-2}
The following equation holds, 
\bea
&&
W_e=H^0(C,L_{\Delta,e}(-\sum_{j=1}^L P_j+\ast p_\infty))\oplus 
\oplus_{m=1}^M {\mathbb C}\varphi_m.
\label{decomp-We}
\ena
\end{lemma}
\noindent
{\it Proof.} It is obvious that the right hand side is included in the left hand side.
Let us prove the converse inclusion.

Take any $f\in W_e$ and set $f(q_i)=c_i$. Set 
\bea
&&
F=f-\sum_{m=1}^M c_m \varphi_m.
\non
\ena
Then 
\bea
&&
F(q_i)=f(q_i)-\sum_{m=1}^M c_m \varphi_m(q_i)=c_i-c_i=0.
\non
\ena
Since $f$ and $\varphi_l$ are both in $W_e$, $F\in W_e$ and $F(p_i)=0$ for 
$1\leq i\leq N$. Therefore 
\bea
&&
F\in H^0(C,L_{\Delta,e}(-\sum_{j=1}^L P_j+\ast p_\infty))
\non
\ena
and therefore $f$ is in the right hand side of (\ref{decomp-We}).
\qed
\vskip5mm

Take $n$ larger than the order of a pole of $\varphi_j$ at $p_\infty$ for any $j$.
Then 
\bea
&&
W_e(n)=H^0(C,L_{\Delta,e}(-\sum_{j=1}^L P_j+n p_\infty))\oplus 
\oplus_{m=1}^M {\mathbb C}\varphi_m,
\non
\ena
and
\bea
&&
\dim W_e(n)=\dim H^0(C,L_{\Delta,e}(-\sum_{j=1}^L P_j+n p_\infty)) + M.
\non
\ena
 
If we further take $n$ larger than $g-1+L$, 
\bea
&&
\deg L_{\Delta,e}(-\sum_{j=1}^L P_j+n p_\infty)=g-1-L+n>2g-2=\deg \Omega^1,
\non
\ena
then
\bea
&&
H^1(C, L_{\Delta,e}(-\sum_{j=1}^L P_j+n p_\infty))=0,
\non
\ena
and, by the Riemann-Roch theorem, 
\bea
&&
\dim H^0(C,L_{\Delta,e}(-\sum_{j=1}^L P_j+n p_\infty))=n-L.
\non
\ena
Therefore, for $n>>0$, 
\bea
&&
\dim W_e(n)=n-L+M=n-N.
\non
\ena
\qed

\noindent
{\bf Proof of Lemma \ref{lemma-1}.} \par
\vskip2mm

We first show that, for each $1\leq i\leq L$, there exists an element $G_i$ of 
$H^0(C, L_{\Delta,e}(\ast p_\infty))$ such that it has only a simple pole 
at $P_i$ on $C\backslash\{p_\infty\}$.

By the Riemann-Roch theorem, for a sufficiently large $n$, we have
\bea
&&
\dim H^0(C, L_{\Delta,e}(P_i+n p_\infty))=n+1,
\non
\\
&&
\dim H^0(C, L_{\Delta,e}(n p_\infty))=n.
\non
\ena
It means that there exists a meromorphic section of $L_{\Delta,e}$ which 
has a simple pole at $P_i$, a pole at $p_\infty$ of order at most $n$ and 
has no other poles. Thus $G_i$ exists.

Next we prove that there exists a meromorphic function on $C$ such that it has a simple zero at every $P_i$, $1\leq i\leq L$, and it is holomorphic on $C\backslash\{p_\infty\}$.

Again, if we take $n$ sufficiently large, we have, by the Riemann-Roch theorem,
\bea
&&
\dim H^0(C, {\cal O}(-\sum_{j=1}^L P_j+n p_\infty))=1-g-L+n,
\non
\\
&&
\dim H^0(C, {\cal O}(-P_i-\sum_{j=1}^L P_j+n p_\infty))=-g-L+n.
\non
\ena
It follows that, for each $i\leq L$, there exists a meromorphic function $h_i$
on $C$ which has a simple zero at $P_i$, a zero at $P_j$, $j\neq i$, and 
is holomorphic on $C\backslash\{p_\infty\}$.
We shall show that a desired $h$ can be constructed as a linear combination of 
$\{h_i\}$.

Set
\bea
&&
h=\sum_{i=1}^L \lambda_i h_i.
\non
\ena

At each $P_i$ take a local coordinate $w$ and write
\bea
&&
h_i=c_i w+O(w^2),
\hskip10mm
h_j=c_j w^{m_j}+O(w^{m_j+1}), \quad j\neq i,
\non
\ena
where $c_j\neq 0$ for any $j$.
Let $\{j\, |\, m_j=1\}=\{j_1,...,j_r\}$, where we set $m_i=1$.
Then 
\bea
&&
h=(c_{j_1}\lambda_{j_1}+\cdots+c_{j_r}\lambda_{j_r})w+O(w^2),
\non
\ena
and the condition that $h$ has a simple zero at $P_i$ is 
\bea
&&
c_{j_1}\lambda_{j_1}+\cdots+c_{j_r}\lambda_{j_r}\neq0.
\label{genericity}
\ena
This is an open condition for $\{\lambda_j\}$. Thus there exists $\{\lambda_j\}$
such that $h$ has a simple zero at every $P_i$.

Choosing one set of  $\{h,G_i |1\leq i\leq L\}$ define the element of $H^0(C,L_{\Delta,e}(\ast p_\infty))$ by
\bea
&&
F_i=a_i h G_i, 
\quad
a_i=(hG_i)(p_i)^{-1}.
\non
\ena
Obviously they satisfy $F_i(p_j)=\delta_{i,j}$.
\qed

\section{Proof of Theorem \ref{main} (ii)}
Let $\Psi(t,z)$ be the wave function of $\tau(t)$,
\bea
&&
\Psi(t,z)=\frac{\tau(t-[z])}{\tau(t)} e^{\eta(t,z^{-1})}.
\label{def-Psi}
\ena
We first show 

\begin{lemma}
For $1\leq i\leq N$ 
\bea
&&
\Psi(t,p_i)=-\sum_{j=1}^M \tb_{N+j,i}\Psi(t,p_{N+j}), 
\quad
\tb_{N+j,i}=b_{N+j,i}(p_ip_{N+j}^{-1})^N
\label{eq-Psi}
\ena
\end{lemma}
\vskip2mm
\noindent
{\it Proof.}
Substitute (\ref{def-Psi}) into (\ref{eq-Psi}) we have the equation for
$\tau(t)$,
\bea
&&
\tau(t-[p_i])e^{\eta_i}=-\sum_{j=1}^M \tb_{N+j,i}\tau(t-[p_{N+j}])e^{\eta_{N+j}}.
\label{eq-tau}
\ena
Let us prove this equation.

 By (\ref{tau-explicit}) we have, using $e^{-\eta([z],p^{-1})}=1-p^{-1}z$,
 \bea
 &&
 \tau(t-[z])=
 \sum_I B_I \Delta^{-}_I \prod_{j\in I}(1-p_j^{-1}z)\, e^{\eta_I}\tau_0(t-[z]-[p_I]).
 \label{tau-tz}
 \ena
Substituting $z=p_i$, $1\leq i\leq N$ and multiplying by $e^{\eta_i}$ we get
 \bea
 \tau(t-[p_i])e^{\eta_i}
 &=&
 p_i^N\sum_I B_I\Delta_I^{-}\prod_{j\in I}(p_i^{-1}-p_j^{-1})
 e^{{\eta_I}+\eta_i}\tau_0(t-[p_i]-[p_I])
 \non
 \\
 &=&
 p_i^N\sum_{i\notin I} B_I\Delta_{(I,i)}^{-}
 e^{\eta_{(I,i)}}\tau_0(t-[p_{(I,i)}]).
 \label{tau-pi-0}
 \ena
 
 We write $I=(I',N+j_1,...,N+j_r)$ with 
 \bea
 &&
 I'\in \binom{[N]}{N-r}, 
 \quad
 1\leq j_1<\cdots<j_r\leq M.
 \non
 \ena
 Since $i\notin I$, we have $r\geq 1$ and $i\notin I'$. 
 For simplicity we set
 \bea
 &&
 T(I,i)= e^{\eta_{(I,i)}}\tau_0(t-[p_{(I,i)}]).
 \label{def-T}
 \ena
 Then
 \bea
 &&
 \text{RHS of (\ref{tau-pi-0})}
 \non
 \\
 &=&
 p_i^N\sum_{r=1}^N \sum_{ I'\in \binom{[N]}{N-r}, i\notin I', 1\leq j_1<\cdots<j_r\leq M}
 B_{(I',N+j_1,...,N+j_r)}\Delta^-_{(I',N+j_1,...,N+j_r,i)}
 \non
 \\
 &&
 \times
 T(I',N+j_1,...,N+j_r,i).
 \label{tau-pi-1}
 \ena
 To proceed we extend the index $I$ of $B_I$ and $\Delta^{-}_I$ to 
 arbitrary sequence from $[L]$ in a skew symmetric way.  In particular,
 for $I=(i_1,...,i_N)$, $B_I\Delta^{-}_I$ is symmetric in $i_1,...,i_N$ and $B_I$, $\Delta^-_I$ become $0$ if some of indices in $I$ coincide.
 
 Then
 \bea
 &&
 \sum_{1\leq j_1<\cdots<j_r\leq M}B_{(I',N+j_1,...,N+j_r)}\Delta^{-}_{(I',N+j_1,...,N+j_r,i)}
 T(I',N+j_1,...,N+j_r,i)
 \non
 \\
 &=&
 \frac{1}{r!}\sum_{j_1,...,j_r=1}^M
 B_{(I',N+j_1,...,N+j_r)}\Delta^{-}_{(I',N+j_1,...,N+j_r,i)}
 T(I',N+j_1,...,N+j_r,i).
 \label{tau-pi-2}
 \ena
 Recall that $B$ has the form
 \bea
 &&
 B=\left(
 \begin{array}{ccc}
 1&\quad&\quad\\
 \quad&\ddots&\quad\\
 \quad&\quad&1\\
 b_{N+1,1}&\cdots&b_{N+1,N}\\
 \vdots&\quad&\vdots\\
 b_{N+M,1}&\cdots&b_{N+M,N}\\
 \end{array}
 \right).
 \non
 \ena
 and $i\notin I'$. 
 Then the expansion of the determinant $B_{(I',N+j_1,...,N+j_r)}$ in the 
 $i$-th column takes the form
 \bea
 &&
 B_{(I',N+j_1,...,N+j_r)}
 =
 \sum_{k=1}^r (-1)^{N-r+k+i}b_{N+j_k,i}
 B^{(i)}_{(I',N+j_1,...,\widehat{N+j_k},...,N+j_r)}.
 \label{tau-pi-3}
 \ena
 Here $B^{(i)}_{i_1,...,i_{N-1}}$ denotes the determinant 
 $\det(b_{i_m,j})_{1\leq m\leq N-1,1\leq j\leq N,j\neq i}$.
 Substitute (\ref{tau-pi-3}) into (\ref{tau-pi-2}) and change the order of sum:
 \bea
&&
 \text{RHS of (\ref{tau-pi-2})}
 \non
 \\
 &=&
 \frac{1}{r!}\sum_{k=1}^r
 \sum_{j_1,...,j_r=1}^M
 (-1)^{N-r+k+i}b_{N+j_k,i}
 B^{(i)}_{(I',N+j_1,...,\widehat{N+j_k},...,N+j_r)}
 \Delta^{-}_{(I',N+j_1,...,N+j_r,i)}
 \non
 \\
 &&
 \times
 T(I',N+j_1,...,N+j_r,i).
 \label{tau-pi-4}
 \ena
 Use 
 \bea
 &&
 \Delta^{-}_{(I',N+j_1,...,N+j_r,i)}=(-1)^{r-k}
 \Delta^{-}_{(I',N+j_1,...,\widehat{N+j_k},...,N+j_r,N+j_k,i)}
 \non
 \ena
 and change the names of indices as $j_k\rightarrow j$, $j_1,...,\widehat{j_k},...,j_r$
 $\rightarrow$ $j_1,...,j_{r-1}$. We get
 \bea
 &&
 \text{RHS of (\ref{tau-pi-4})}
 \non
 \\
 &=&
 \frac{1}{r!}\sum_{k=1}^r
 (-1)^{N+i}\sum_{j=1}^M b_{N+j,i}\sum_{j_1,...,j_{r-1}=1}^M
 B^{(i)}_{(I',N+j_1,...,N+j_{r-1})}
 \Delta^{-}_{(I',N+j_1,...,N+j_{r-1},N+j,i)}
 \non
 \\
 &&
 \times
 T_{(I',N+j_1,...,N+j_{r-1},N+j,i)}.
 \non
 \ena
 Since the summand does not depend on $k$, the sum in $k$ gives
 $r$ times of the summand. 
 Rewrite the summation in $1\leq j_1,...,j_{r-1}\leq M$ to $(r-1)!$ times 
 the summation in $(j_1,...,j_{r-1})$ with $1< j_1<\cdots<j_{r-1}\leq M$.
 Subsitute it into (\ref{tau-pi-1}) and get
 \bea
 &&
 \text{RHS of (\ref{tau-pi-1})}
 \non
 \\
 &=&
 (-1)^{N+i}p_i^N\sum_{j=1}^Nb_{N+j,i}\sum_{r=1}^M\sum_{I'\in\binom{[N]}{N-r},i\notin I'}
 \sum_{1\leq j_1<\cdots<j_{r-1}\leq M}
 B^{(i)}_{(I',N+j_1,...,N+j_{r-1})}
 \non
 \\
 &&
 \times
 \Delta^{-}_{(I',N+j_1,...,N+j_{r-1},N+j,i)}
 T_{(I',N+j_1,...,N+j_{r-1},N+j,i)}.
 \non
 \ena
 Notice that taking the summation over $r$, $I'$, $\{j_k\}$ is equivalent to taking the summation over 
 $I'\in\binom{[L]}{N-1}$ with $i,N+j \notin I'$. Thus 
 \bea
 &&
\tau(t-[p_i])e^{\eta_i}
\non
\\
&= &\text{RHS of (\ref{tau-pi-1})}
 \non
 \\
 &=&
 (-1)^{N+i}p_i^N\sum_{j=1}^Nb_{N+j,i}\sum_{I'\in\binom{[L]}{N-1}, i,N+j \notin I'}
 B^{(i)}_{I'}\Delta^{-}_{(I',N+j,i)}T(I',N+j,i).
 \label{tau-pi-5}
 \ena
 
 Next, by replacing $i$ by $N+j$ in (\ref{tau-pi-0}) and recalling 
 the definition (\ref{def-T}) of $T(I,i)$,  we have 
 \bea
 \tau(t-[p_{N+j}])e^{\eta_{N+j}}
 =
 p_{N+j}^N\sum_{N+j\notin I} B_I\Delta_{(I,N+j)}^{-}
 T(I,N+j).
 \ena
 It follows that
 \bea
 &&
 \sum_{j=1}^M \tb_{N+j,i} \tau(t-[p_{N+j}])e^{\eta_{N+j}}
 \non
 \\
 &=&
 p_{i}^N \sum_{j=1}^M b_{N+j,i}\sum_{N+j\notin I} B_I\Delta_{(I,N+j)}^{-}
 T(I,N+j)
 \label{tau-pi-6}
 \\
 &=&
 I_{+}+I_{-},
 \label{tau-pi-7}
 \ena
 where $I_+$ is the part of the RHS of (\ref{tau-pi-6}) such that 
 $I$ includes $i$ in the summation over $I$ and $I_{-}$ the part where
 $I$ does not include $i$.
 
 We show that $I_{+}$ is equal to $-\tau(t-[p_{i}])e^{\eta_{i}}$
 and $I_{-}=0$.
 
 Let us first consider $I_{+}$.
 Separating $i$ from $I$ we have
 \bea
 &&
 I_{+}=p_i^N\sum_{j=1}^M b_{N+j,i}
 \sum_{I'\in\binom{[L]}{N-1}, i,N+j\notin I} B_{(I',i)}\Delta_{(I',i,N+j)}^{-}
 T(I',i,N+j).
 \non
 \ena
 Since the $N$-th row vector of $B_{(I',i)}$ is the $i$-th unit vector, we have
 \bea
 &&
 B_{(I',i)}=(-1)^{N+i}B_{I'}^{(i)}.
 \non
 \ena
 Therefore
 \bea
  I_{+}&=&(-1)^{N+i}p_i^N\sum_{j=1}^M b_{N+j,i}
 \sum_{I'\in\binom{[L]}{N-1}, i,N+j\notin I} B_{I'}^{(i)}\Delta_{(I',i,N+j)}^{-}
 T(I',i,N+j)
 \non
 \\
 &=&
 -\text{RHS of (\ref{tau-pi-5})}
 \non
 \\
 &=&
 -\tau(t-[p_{i}])e^{\eta_{i}},
 \label{tau-qj-0}
 \ena
 where $\Delta_{(I',i,N+j)}^{-}=-\Delta_{(I',N+j,i)}^{-}$ is used.
 
 Next let us consider $I_{-}$. In a similar computation to deriving (\ref{tau-pi-5})
 we have
 \bea
 I_{-}&=&
 (-1)^{N+i}\sum_{j=1}^Mb_{N+j,i}\sum_{j'=1}^Mb_{N+j',i}
 \sum_{I'\in\binom{[L]}{N-1}, i,N+j',N+j\notin I'}
 B^{(i)}_{I'}\Delta^{-}_{(I',N+j',N+j)}
 \non
 \\
 &&
 \times 
 T(I',N+j',N+j)
 \non
 \\
 &=&
 (-1)^{N+i}\sum_{j,j'=1}^Mb_{N+j,i}b_{N+j',i}
 \sum_{I'\in\binom{[L]}{N-1}, i,N+j',N+j\notin I'}
 B^{(i)}_{I'}\Delta^{-}_{(I',N+j',N+j)}
 \non
 \\
 &&
 \times 
 T(I',N+j',N+j).
 \non
 \ena
 Since $b_{N+j,i}b_{N+j',i}$ is symmetric in $j, j'$ and the remaining part is 
 skew symmetric in $j, j'$, the last summation in $j, j'$ becomes zero. Therefore
 $I_{-}=0$. We, then, have (\ref{eq-tau}) by  (\ref{tau-pi-7}), (\ref{tau-qj-0}). \qed
 
 By Lemma \ref{exp-Q} we have
 
 \begin{lemma}\label{lem-3}
 Let $N\geq 1$, $Q_j\in C$, $1\leq j\leq N$ and $z_j=z(Q_j)$.
 Then
 \bea
 &&
 \tau_0(t-[z]-\sum_{j=1}^N[z_j])e^{\eta(t,z^{-1})}
 \non
 \\
 &=&
 \left(\frac{z}{E(0,z)}\right)^{N+1}
 \prod_{j=1}^N\frac{E(z,z_j)}{z_j-z}
 \prod_{j=1}^N \frac{z_j}{E(0,z_j)}
 e^{q(\sum_{j=1}^N[z_j])-\sum_{j=1}^N Q(t|[z_j])+\frac{1}{2}q(t)}
 \non
 \\
 &&
 \times
 \theta(\CV t-I(p)-\sum_{j=1}^NI(Q_j)+e)
 e^{\sum_{j=1}^\infty t_j \int^p d\tilde{r}_j}.
 \label{psi-z-1}
 \ena
 \end{lemma}
 
 We use (\ref{psi-z-1}) to compute (\ref{tau-tz}) and get
 \bea
 &&
 \tau(t-[z])e^{\eta(t,z^{-1})}
 \non
 \\
 &=&
z^{N+1} \sum_{J\in \binom{[L]}{N}}
B_J\Delta^{-}_J e^{\eta_J}
\prod_{j\in J}
\frac{E(z,p_j)}{E(0,z)E(0,p_j)}
e^{q(\sum_{j\in J}[p_j])-\sum_{j\in J}Q(t|[p_j])+\frac{1}{2}q(t)}
\non
\\
&&
\times
\frac{1}{E(0,z)}
  \theta(\CV t-I(p)-\sum_{j\in J}I(Q_j)+e)
 e^{\sum_{j=1}^\infty t_j \int^p d\tilde{r}_j}.
 \label{psi-z-2}
 \ena
 Set
 \bea
 &&
 \Psi'(t,z)=z^{-N-1}\Psi(t,z).
 \non
 \ena
 Then (\ref{psi-z-2}) shows that 
 the expansion coefficients of 
 $\tau(t)\Psi'(t,z)=z^{-N-1}\tau(t-[z])e^{\eta(t,z^{-1})}$ belong to 
 $H^0(C, L_{\Delta,e}(\ast p_\infty))$.
 Rewriting the equation (\ref{eq-Psi}) in terms of $\Psi'(t,z)$ we have
 \bea
 &&
 \Psi'(t,p_i)=-\sum_{j=1}^M b'_{N+j,i} \Psi'(t,q_j).
 \non
 \ena
 It means that the expansion coefficients of $\tau(t)\Psi'(t,z)$ are in $W_{e}$.
 Thus the expansion coefficients of $\tau(t)\Psi(t,z)$ are in $z^{N+1}W_{e}=U_e$.
 Since $U_e\in UGM$ by (1) of Theorem \ref{main} and the strict inclusion relation is impossible for points of UGM, $U_e$ is the point of UGM corresponding to $\tau(-t)$.
 \qed

\appendix

\section{Proof of Proposition \ref{stabilizer}}

We first show that $R_e$ is contained in the RHS of (\ref{Re}).
Let $f\in R_e$ and $\varphi_m\in W_e$ be defined in (\ref{phi-m}).
Then $f\varphi_m\in W_e$. 
By (\ref{phi-m-value}) we see that
\bea
&&
(f\varphi_m)(p_i)=-\sum_{j=1}^M b'_{N+j,i} (f\varphi_m)(q_j)
\non
\ena
is equivalent to
\bea
&&
f(p_i)b'_{N+m,i}=b'_{N+m,i}f(q_m).
\label{eq-fb} 
\ena
Therefore $f$ is contained in the RHS of (\ref{Re}).

Let us prove the converse inclusion. Let $f$ be an element of the RHS of
(\ref{Re}) and $F\in W_e$.  Notice that the equation (\ref{eq-fb}) holds for any $i,m$. 
Then
\bea
-\sum_{j=1}^M b'_{N+j,i} (fF)(q_j)
&=&
-\sum_{j=1}^M b'_{N+j,i} f(q_j)F(q_j)
\non
\\
&=&
-\sum_{j=1}^M b'_{N+j,i} f(p_i)F(q_j)
\non
\\
&=&
f(p_i)\left(-\sum_{j=1}^M b'_{N+j,i} F(q_j)\right)
\non
\\
&=&
f(p_i)F(p_i)=(fF)(p_i)
\non
\ena
which means $fF\in W_e$. Thus $f\in R_e$. 
\qed

\section{Proof of Theorem \ref{C'-structure}}

(iii) follows from (ii). Let us prove (i), (ii), (iv).

\noindent
 (i) It can be easily proved that if $\varphi({\cal P})=p'_\infty$ for some ${\cal P}\in 
\Spec R_e$ then ${\cal P}=R_e$.  It is absurd.

\noindent
(ii) 
Let ${\cal P}=\oplus_{n=0}^\infty {\cal P}^{(n)}\in C'$ such that ${\cal P}'\neq p'_\infty$. Set 
\bea
&&
q=\cup_{n=0}^\infty {\cal P}^{(n)}\subset Re,
\non
\ena
which obviously becomes an ideal of $R_e$.
It is sufficient to prove the following lemma.

\begin{lemma}
(i) If $x,y\in R_e$ satisfies $xy\in q$, either $x\in q$ or $y\in q$.
\vskip2mm
\noindent
(ii) $q\neq R_e$.
\vskip2mm
\noindent
(iii) $\varphi(q)={\cal P}$.
\end{lemma}
\noindent
{\it Proof.} 
(i) We can assume that
\bea
&&
x=z^{-m}+O(z^{-m+1}),
\hskip10mm
y=z^{-n}+O(z^{-n+1})
\non
\ena
with $m,n\geq 1$. Then 
\bea
&&
xy\in R_e(m+n)\backslash R_e(m+n-1).
\label{lem-1}
\ena
Since $xy\in q$, there exists $N\geq 0$ such that 
\bea
&&
xy\in {\cal P}^{(N)}\subset R_e(N).
\non
\ena
By (\ref{lem-1}) we have $N\geq m+n$.
Set 
\bea
&&
N-(m+n)=k\geq 0.
\non
\ena
Then 
\bea
&&
x\in R_e(m)\subset R_e(m+k),
\hskip10mm
y\in R_e(n).
\non
\ena
So we consider $x$ and $y$ as homogeneous elements of $A'$ with degree 
$m+k$ and $n$ respectively. Then 
\bea
&&
xy\in {\cal P}^{(N)}\subset A'.
\non
\ena
Since ${\cal P}$ is a prime ideal of $A'$, $x\in {\cal P}$ or $y\in {\cal P}$.
Therefore $x\in {\cal P}\cap R_e(m+k)= {\cal P}^{(m+k)}$ or $y\in {\cal P}\cap 
R_e(m)={\cal P}^{(m)}$. It means that $x\in q$ or $y\in q$. 
\vskip2mm

\noindent
(ii) Notice that $1\in R_e(n)$ for any $n\geq 0$. If we consider $1$ as a homogeneous
element of $A'$ with degree $n$ we denote it by $1^{(n)}$. We shall show 
\bea
&&
1^{(n)}\notin {\cal P}^{(n)},
\hskip10mm
n\geq 0.
\label{lem-2}
\ena
Since ${\cal P}\neq A'$, $1^{(0)}\notin {\cal P}^{(0)}$.

Let us consider the case $n=1$. Suppose that $1^{(1)}\in {\cal P}^{(1)}$.
Notice that $A'1^{(1)}=p'_\infty$. Therefore 
\bea
&&
p'_\infty\subsetneq {\cal P},
\non
\ena
since ${\cal P}\neq p'_\infty$. Then there exists $k\geq 1$ and $f_k\in {\cal P}^{(k)}$
such that $f_k\notin R_e(k-1)$, that is, 
\bea
&&
f_k=z^{-k}+O(z^{-k+1}).
\non
\ena
By the Riemann-Roch theorem we have 
\bea
&&
\dim R_e(n)/R_e(n-1)=1.
\hskip10mm 
n>>0,
\ena

It follows that, for all sufficiently large $n$, there exists $h_n\in R_e(n)$ such that
\bea
&&
h_n=z^{-n}+O(z^{-n+1}).
\non
\ena
Then $h_nf_k\in {\cal P}^{(n+k)}$ and $h_nf_k\in R_e(n+k)\backslash R_e(n+k-1)$.
Taking into account that ${\cal P}^{(n+k)}\supset R_e(n+k-1)$ we see that
\bea
&&
{\cal P}^{(n)}=R_e(n), 
\hskip10mm
n>>0.
\label{lem-3}
\ena

On the other hand 
\bea
&&
{\cal P}\nsupseteq \oplus_{n=1}^\infty R_e(n)
\non
\ena
since ${\cal P}\in C'$.
It follows that there exists $N\geq 1$ and $F_N\in R_e(N)$ such that
$F_N\notin {\cal P}^{(N)}$.
Since ${\cal P}^{(N)}\supset R_e(N-1)$, $F_N\in R_e(N)\backslash R_e(N-1)$,
that is, 
\bea
&&
F_N=z^{-N}+O(z^{-N+1}).
\non
\ena
Since ${\cal P}$ is a prime ideal,
\bea
&&
F_N^m\notin {\cal P}^{(Nm)},
\non
\ena
for any $m\geq 1$ which contradicts (\ref{lem-3}). 
Thus $1^{(1)}\notin {\cal P}^{(1)}$.

Notice that $(1^{(1)})^n=1^{(n)}$ for $n\geq 2$. Thus $1^{(n)}\notin {\cal P}^{(n)}$ and (\ref{lem-2}) has been proved. 

Then $1\notin q$ and $q\neq R_e$.
\vskip2mm

\noindent
(iii) We have to prove that $q\cap R_e(n)={\cal P}^{(n)}$ for any $n\geq 0$.
Suppose that this does not hold. Then there exists $N\geq 1$ such that 
\bea
&&
q\cap R_e(N)\neq {\cal P}^{(N)}.
\non
\ena
Since the right hand side is contained in the left hand side, it means 
\bea
&&
q\cap R_e(N)\supsetneq{\cal P}^{(N)}.
\non
\ena
Therefore there exists $x\in q\cap R_e(N)$ such that $x\notin {\cal P}^{(N)}$.
Since $x\in q$ there exists $M$ such that $x\in {\cal P}^{(M)}$. Then 
$N<M$. In fact if $N\geq M$, $x\in {\cal P}^{(M)}\subset {\cal P}^{(N)}$.
Since $x\in q\cap R_e(N)$, we can consider $x$ as a homogeneous element of $A'$ with degree $N$. Then $x\notin {\cal P}^{(N)}$ and $1^{(M-N)}\notin {\cal P}^{(M-N)}$ but
\bea
&&
1^{(M-N)} x=x\in {\cal P}^{(M)},
\non
\ena which is absurd. Thus $q\cap R_e(n)={\cal P}^{(n)}$ for any $n$ and 
(iii) of the lemma is proved.
\qed

\vskip5mm
\noindent
Proof of  Theorem \ref{C'-structure} (iv).
\vskip2mm
It is obvious that $a\notin p'_\infty$ and $p'_\infty\in D_{+}(a)$.
Let 
\bea
&&
(p'_\infty)_{(a)}=\sum_{n=0}^\infty \frac{R_e(nm-1)}{a^n}
\non
\ena
be the image of $p'_\infty$ in $A'_{(a)}$. Any element of $f\in R_e(nm)$ satisfies 
$f-ca^n\in R_e(nm-1)$ for some constant $c\in {\mathbb C}$. It means that $f/a^n=c$ modulo $(p_\infty)_{(a)}$. Since $(p_\infty)_{(a)}\neq A_{(a)}$, this means
that $A_{(a)}/(p'_\infty)_{(a)}\simeq {\mathbb C}$. Thus $(p'_\infty)_{(a)}$ is a maximal ideal of $A_{(a)}$ and  (iv) is proved,
\qed

\section{Proof of Proposition \ref{torsion-free}}

 Both $W_e$ and $R_e$ is a subspace of ${\mathbb C}((z))$ and the $R_e$ module 
structure of $W_e$ is given by the ring structure of ${\mathbb C}((z))$. Therefore $W_e$ 
is a torsion free $R_e$ module. That ${\cal W}_e$ is a torsion free ${\cal O}_{C'}$ module follows from this.

Let $K'$ be the quotient field of $R_e$. In order to prove that the rank of $W_e$ is one,  it is sufficient to show, by the definition, that 
$\dim_{K'} K'\otimes_{R_e} W_e=1$.

\begin{lemma}\label{K'}
Let $K$ be the quotient field of $R$. Then $K'=K$.
\end{lemma}
\vskip2mm
\noindent
{\it Proof.} Since $R_e\subset R$, $K'\subset K$. Let us prove the converse inclusion.
Take any $f\in K$.  Let the pole divisor of $f$ be 
\bea
&&
m_1R_1+\cdots+m_sR_s +m_\infty p_\infty,
\quad 
m_i>0, \, R_i\in C,\,  R_i\neq p_\infty.
\non
\ena
Take $n$ sufficiently large such that there exists non zero 
$F\in H^0(C, {\cal O}(-\sum_{i=1}^s m_iR_i-\sum_{i,j} P_{i,j}+np_\infty))$.
Then $F\in R_e$ since $F(p_{i,j})=0$ for all $i,j$.
Set $h=fF$. Then 
\bea
&&
h\in H^0(C,{\cal O}(-\sum_{i,j} P_{i,j}+\ast p_\infty))\subset R_e.
\non
\ena
Thus $f=h/F\in K'$. Thus $K\subset K'$.
\qed

Let us continue the proof of Proposition \ref{torsion-free}. 
Take any nonzero $f_1\in W_e$. Notice that $K$ is the field of meromorphic 
functions on $C$. Therefore, for any $f_2\in W_e$, $f_2/f_1\in K$ and $f_2\in Kf_1$.
By Lemma \ref{K'} we have $ W_e\subset K f_1=K'f_1$. Thus $K'\otimes_{R_e}W_e=K'f_1$
and $\dim_{K'} K'\otimes_{R_e}W_e=1$.
\qed

\section{Proof of Proposition \ref{normalization}}
We begin by determining the structure of $R$ and $R_e$. 

\begin{lemma}\label{Hij}
For any $(i,j)$, $1\leq i\leq s$, $1\leq j\leq n_i$, there exists $H_{i,j}\in R$ such that
\bea
&&
H_{i,j}(p_{k,l})=\delta_{i,k}\delta_{j,l}.
\non
\ena
\end{lemma}
\noindent
{\it Proof.} By the Riemann-Roch formula, for all sufficiently large $n$, we have
\bea
&&
\dim H^{0}\left(C,{\cal O}(-\sum P_{k,l}+np_\infty)\right)=n-L+1-g,
\non
\\
&&
\dim H^{0}\biggl(C,{\cal O}(-\sum_{(k,l)\neq (i,j)} P_{k,l}+np_\infty)\biggr)=n-L+2-g.
\non
\ena
A non-zero element of the latter space which does not belong to the former space satisfies the required property if it is adjusted by a constant multiple.
\qed

Let 
\bea
&&
H_i=H_{i,1}+\cdots+H_{i,n_i}.
\non
\ena

Then
\begin{lemma}
(i) $H_i(p_{i,j})=1$ for $1\leq j\leq n_i$ and $H_i(p_{k,l})=0$ if $k\neq i$.
\vskip2mm
\noindent
(ii) $H_i\in R_e$.
\vskip2mm
\noindent
(iii) $\{H_i\, |\, 1\leq i\leq s\}$ is linearly independent.
\end{lemma}
The lemma can be easily proved from the definition of $H_i$. 
So we leave the proof to the reader.

\begin{prop}\label{structure-R-Re}
(i) $R=H^0\left(C, {\cal O}(-\sum P_{i,j}+\ast p_\infty)\right)
\oplus\oplus_{i,j} {\mathbb C} H_{i,j}$.
\vskip2mm
\noindent
(ii) $R_e=H^0\left(C, {\cal O}(-\sum P_{i,j}+\ast p_\infty)\right)\oplus\oplus_{i=1}^s {\mathbb C} H_i$.
\end{prop}
\vskip2mm
\noindent
{\it Proof.} (i) It is sufficient to prove that the left hand side is contained in the 
right hand side. Take any $f\in R$. Set $f(p_{i,j})=c_{i,j}$ and $f'=f-\sum c_{i,j}H_{i,j}$.
Then $f'(p_{i,j})=0$ for any $i,j$. Thus 
\bea
&&
f' \in H^0\left(C, {\cal O}(-\sum P_{i,j}+\ast p_\infty)\right),
\non
\ena
which shows that $f$ is contained in the RHS of (i).

(ii) is similarly proved.
\qed

Since $R_e$ is a subring of $R$, $R$ is considered as an $R_e$-module. 
Then

\begin{corollary}\label{R-finite-Re}
The ring $R$ is a finitely generated $R_e$ module.
\end{corollary}
\vskip2mm
\noindent
{\it Proof.} By Proposition \ref{structure-R-Re} we have
\bea
&&
R=R_e 1+\sum R_e H_{i,j},
\non
\ena
which shows the assertion of the corollary.
\qed

Since $C$ is non-singular $R$ is integrally closed in $K$.

Then we have
\begin{corollary}\label{normal-1}
The ring $R$ is the integral closure of $R_e$ in $K$.
\end{corollary}
\vskip2mm
\noindent
{\it Proof.} By Corollary \ref{R-finite-Re} $R$ is integral over $R_e$.
An integral element of $K$ over $R_e$ is integral over $R$ and therefore
it belongs to $R$ since $R$ is integrally closed. Thus $R$ is the integral closure
in $K$.
\qed

Take $a\in R_e(m)\subset R(m)$ as in (\ref{element-a}).
Consider $a$ as an element of $A'$ and $A$ with the degree $m$.

Similarly to the above corollary we can prove the following.

\begin{prop}\label{normal-2}
The ring $A_{(a)}$ is a finitely generated $A'_{(a)}$ module and it is the integral 
closure of $A'_{(a)}$ in $K$.
\end{prop}

By Corollary \ref{normal-1} and Proposition \ref{normal-2} we have
Proposition \ref{normalization}.

\section{Proof of Proposition \ref{singularity}}
(i) Let
\bea
&&
S=R-m=\{f\in R\,|\, f(p)\neq0\},
\hskip5mm
S'=R_e-m'=\{f\in R_e\,|\, f(p)\neq 0\}.
\non
\ena
Then, $S'\subset S$ and, by definition, $R_m=S^{-1}R$, 
$(R_e)_{m'}={S'}^{-1}R_e$. Therefore $(R_e)_{m'}\subset R_m$. 
Let us prove the converse inclusion.
Take any $f\in R_m$ and write it as 
\bea
&&
f=\frac{F}{G},
\hskip5mm
F\in R, G\in S.
\non
\ena
Notice that, by the Riemann-Roch theorem,  there exists $H\in R$ such that
\bea
&&
H\in H^0\left(C, {\cal O}(-\sum P_{i,j}+np_\infty)\right),
\hskip5mm
H(p)\neq 0,
\label{H}
\ena
if $n$ is sufficiently large. Then $FH, GH\in R_e$, $(GH)(p)\neq 0$ and 
$f=FH/GH\in {S'}^{-1}R_e$. Thus $R_m\subset (R_e)_{m'}$.
\par
\noindent
(ii) Since $R$ is the integral closure of $R_e$, the integral closure of $(R_e)_{m'}={S'}^{-1}R_e$ is ${S'}^{-1}R$ (c.f. Proposition 2.1 of \cite{Iitaka}).
Consider $H_{i,j}\in R\subset {S'}^{-1}R$ of Lemma \ref{Hij}. It is not in ${S'}^{-1}R_e$. In fact if $H_{i,j}\in {S'}^{-1}R_e$ then $fH_{i,j}\in R_e$ for some $f\in S'$. 
Then $(fH_{i,j})(p_{i,j})=f(p_{i,j})\neq 0$ and 
$(fH_{i,j})(p_{i,j'})=f(p_{i,j'})H(p_{i,j'})=0$ for $j'\neq j$ which contradicts $fH_{i,j}\in R_e$.
Thus ${S'}^{-1}R_e\neq {S'}^{-1}R$ and ${S'}^{-1}R_e$ is not a normal ring.
\par
Let us prove the last statement of (ii).
Obviously $m_{P_{i,j'}}\in \psi^{-1}(m')$ for any $j'$.
Since $R$ is integral over $R_e$, each element of $\psi^{-1}(m')$ is a maximal
ideal. 

Let $Q$ be a point of the Riemann surface $C$ such that $Q\neq P_{i,j'}$
for any $j'$ and $z(Q)=q$. 
Suppose that $\psi(m_Q)=m'$. Similarly to (\ref{H}) there exists 
$H\in R_e$ such that $H(p_{i',j'})=0$ for any $i',j'$ and $H(q)\neq 0$. 
Then $H\in m'$ but $H\notin \psi(m_Q)$. It contradicts the assumption. Thus the assertion is proved.
\qed

\vskip10mm
\noindent
{\bf \large Acknowledegements}
\vskip2mm
\noindent
Parts of the results of this paper were presented in a series of lectures at University of Tokyo in July, 2022 and at Nagoya University in July, 2023.
I would like to thank Junichi Shiraishi and Masashi Hamanaka 
for invitations and hospitality. I would also like to thank Hiroaki Kanno,
Yasuhiko Yamada, Shintaro Yanagida for their interests.
I would especially like to thank Yuji Kodama for explaining the contents of the paper \cite{Kodama2023-2} as well as related results \cite{Kodama2023-1} and for his valuable comments when I gave lectures at Nagoya University.
I am also grateful to Yuji Kodama for letting me know his theory on
soliton solutions of the KP equation some years ago. 
That was the starting point of this study.
This work was supported by JSPS KAKENHI Grant Number JP19K03528.

\end{document}